\documentclass[twocolumn]{aastex701}
\usepackage{graphicx}
\usepackage{threeparttable}
\usepackage{booktabs}
\usepackage{longtable}
\usepackage{threeparttablex}
\usepackage{amsmath}
\usepackage[caption=false]{subfig}
\usepackage{comment}
\usepackage{listings}
\usepackage{import}

\lstdefinestyle{json}{
    basicstyle=\ttfamily\small,
    commentstyle=\color{green},
    keywordstyle=\color{blue},
    numberstyle=\tiny\color{gray},
    stringstyle=\color{red},
    breaklines=true,
    captionpos=b,
    showstringspaces=false
}

\newcommand{\figref}[1]{Figure \ref{fig:#1}}
\newcommand{\tabref}[1]{Table \ref{tab:#1}}
\newcommand{\secref}[1]{\S\ref{sec:#1}}

\newcommand{\json}{{\tt json}}

\newcommand{\property}[1]{\textcolor{cyan}{#1}}
\newcommand{\subproperty}[1]{\textcolor{olive}{#1}}

\newcommand{\keyword}[1]{\textcolor{magenta}{#1}}

\newcommand{\ntdes}{240\space}

\begin{document}

\title{The Open mulTiwavelength Transient Event Repository (OTTER): Infrastructure Release and Tidal Disruption Event Catalog}

\newcommand{\LCO}{\affiliation{Las Cumbres Observatory, 6740 Cortona Drive, Suite 102, Goleta, CA 93117-5575, USA}}
\newcommand{\UCSB}{\affiliation{Department of Physics, University of California, Santa Barbara, CA 93106-9530, USA}}
\newcommand{\KITP}{\affiliation{Kavli Institute for Theoretical Physics, University of California, Santa Barbara, CA 93106-4030, USA}}
\newcommand{\UCD}{\affiliation{Department of Physics, University of California, 1 Shields Avenue, Davis, CA 95616-5270, USA}}
\newcommand{\WIS}{\affiliation{Department of Particle Physics and Astrophysics, Weizmann Institute of Science, 76100 Rehovot, Israel}}
\newcommand{\OKC}{\affiliation{Oskar Klein Centre, Department of Astronomy, Stockholm University, Albanova University Centre, SE-106 91 Stockholm, Sweden}}
\newcommand{\OAPD}{\affiliation{INAF-Osservatorio Astronomico di Padova, Vicolo dell'Osservatorio 5, I-35122 Padova, Italy}}
\newcommand{\Caltech}{\affiliation{Cahill Center for Astronomy and Astrophysics, California Institute of Technology, Mail Code 249-17, Pasadena, CA 91125, USA}}
\newcommand{\GSFC}{\affiliation{Astrophysics Science Division, NASA Goddard Space Flight Center, Mail Code 661, Greenbelt, MD 20771, USA}}
\newcommand{\UMD}{\affiliation{Joint Space-Science Institute, University of Maryland, College Park, MD 20742, USA}}
\newcommand{\UCB}{\affiliation{Department of Astronomy, University of California, Berkeley, CA 94720-3411, USA}}
\newcommand{\TTU}{\affiliation{Department of Physics, Texas Tech University, Box 41051, Lubbock, TX 79409-1051, USA}}
\newcommand{\STScI}{\affiliation{Space Telescope Science Institute, 3700 San Martin Drive, Baltimore, MD 21218, USA}}
\newcommand{\UT}{\affiliation{Department of Astronomy, The University of Texas at Austin, 2515 Speedway, Stop C1400, Austin, TX 78712, USA}}
\newcommand{\IoA}{\affiliation{Institute of Astronomy, University of Cambridge, Madingley Road, Cambridge CB3 0HA, UK}}
\newcommand{\QUB}{\affiliation{Astrophysics Research Centre, School of Mathematics and Physics, Queen's University Belfast, Belfast BT7 1NN, UK}}
\newcommand{\IPAC}{\affiliation{Spitzer Science Center, California Institute of Technology, Pasadena, CA 91125, USA}}
\newcommand{\JPL}{\affiliation{Jet Propulsion Laboratory, California Institute of Technology, 4800 Oak Grove Dr, Pasadena, CA 91109, USA}}
\newcommand{\Southampton}{\affiliation{Department of Physics and Astronomy, University of Southampton, Southampton SO17 1BJ, UK}}
\newcommand{\LANL}{\affiliation{Space and Remote Sensing, MS B244, Los Alamos National Laboratory, Los Alamos, NM 87545, USA}}
\newcommand{\Tsinghua}{\affiliation{Physics Department and Tsinghua Center for Astrophysics, Tsinghua University, Beijing, 100084, People's Republic of China}}
\newcommand{\NAOC}{\affiliation{National Astronomical Observatory of China, Chinese Academy of Sciences, Beijing, 100012, People's Republic of China}}
\newcommand{\Itagaki}{\affiliation{Itagaki Astronomical Observatory, Yamagata 990-2492, Japan}}
\newcommand{\Einstein}{\altaffiliation{Einstein Fellow}}
\newcommand{\Hubble}{\altaffiliation{Hubble Fellow}}
\newcommand{\CfA}{\affiliation{Center for Astrophysics \textbar{} Harvard \& Smithsonian, 60 Garden Street, Cambridge, MA 02138-1516, USA}}
\newcommand{\UA}{\affiliation{Department of Astronomy and Steward Observatory, University of Arizona, 933 North Cherry Avenue, Tucson, AZ 85721-0065, USA}}
\newcommand{\MPA}{\affiliation{Max-Planck-Institut f\"ur Astrophysik, Karl-Schwarzschild-Stra\ss e 1, D-85748 Garching, Germany}}
\newcommand{\DSFP}{\altaffiliation{LSSTC Data Science Fellow}}
\newcommand{\HCO}{\affiliation{Harvard College Observatory, 60 Garden Street, Cambridge, MA 02138-1516, USA}}
\newcommand{\Carnegie}{\affiliation{Observatories of the Carnegie Institute for Science, 813 Santa Barbara Street, Pasadena, CA 91101-1232, USA}}
\newcommand{\TAU}{\affiliation{School of Physics and Astronomy, Tel Aviv University, Tel Aviv 69978, Israel}}
\newcommand{\Edinburgh}{\affiliation{Institute for Astronomy, University of Edinburgh, Royal Observatory, Blackford Hill EH9 3HJ, UK}}
\newcommand{\Birmingham}{\affiliation{Birmingham Institute for Gravitational Wave Astronomy and School of Physics and Astronomy, University of Birmingham, Birmingham B15 2TT, UK}}
\newcommand{\CIERA}{\affiliation{Center for Interdisciplinary Exploration and Research in Astrophysics and Department of Physics and Astronomy, \\Northwestern University, 1800 Sherman Ave., 8th Floor, Evanston, IL 60201, USA}}
\newcommand{\Bath}{\affiliation{Department of Physics, University of Bath, Claverton Down, Bath BA2 7AY, UK}}
\newcommand{\CTIO}{\affiliation{Cerro Tololo Inter-American Observatory, National Optical Astronomy Observatory, Casilla 603, La Serena, Chile}}
\newcommand{\Potsdam}{\affiliation{Institut f\"ur Physik und Astronomie, Universit\"at Potsdam, Haus 28, Karl-Liebknecht-Str. 24/25, D-14476 Potsdam-Golm, Germany}}
\newcommand{\INPE}{\affiliation{Instituto Nacional de Pesquisas Espaciais, Avenida dos Astronautas 1758, 12227-010, S\~ao Jos\'e dos Campos -- SP, Brazil}}
\newcommand{\UNC}{\affiliation{Department of Physics and Astronomy, University of North Carolina, 120 East Cameron Avenue, Chapel Hill, NC 27599, USA}}
\newcommand{\Ohio}{\affiliation{Astrophysical Institute, Department of Physics and Astronomy, 251B Clippinger Lab, Ohio University, Athens, OH 45701-2942, USA}}
\newcommand{\AAS}{\affiliation{American Astronomical Society, 1667 K~Street NW, Suite 800, Washington, DC 20006-1681, USA}}
\newcommand{\MMT}{\affiliation{MMT and Steward Observatories, University of Arizona, 933 North Cherry Avenue, Tucson, AZ 85721-0065, USA}}
\newcommand{\Geneva}{\affiliation{ISDC, Department of Astronomy, University of Geneva, Chemin d'\'Ecogia, 16 CH-1290 Versoix, Switzerland}}
\newcommand{\Steward}{\affiliation{Steward Observatory, University of Arizona, 933 North Cherry Avenue, Tucson, AZ 85721, USA}}
\newcommand{\Leiden}{\affiliation{Leiden Observatory, Leiden University, PO Box 9513, 2300 RA Leiden, The Netherlands}}
\newcommand{\PSU}{\affiliation{Department of Astronomy \& Astrophysics, The Pennsylvania State University, University Park, PA 16802, USA}}
\newcommand{\PSUa}{\affiliation{Department of Astronomy \& Astrophysics, The Pennsylvania State University, University Park, PA 16802, USA}}
\newcommand{\PSUb}{\affiliation{Institute for Computational \& Data Sciences, The Pennsylvania State University, University Park, PA 16802, USA}}
\newcommand{\PSUc}{\affiliation{Institute for Gravitation and the Cosmos, The Pennsylvania State University, University Park, PA 16802, USA}}
\newcommand{\IAIFI}{\affiliation{The NSF AI Institute for Artificial Intelligence and Fundamental Interactions, USA}}
\newcommand{\JHU}{\affiliation{Department of Physics and Astronomy, Johns Hopkins University, 3400 North Charles Street, Baltimore, MD 21218, USA}}
\newcommand{\Utah}{\affiliation{Department of Physics \& Astronomy, University of Utah, Salt Lake City, UT 84112, USA}}
\newcommand{\UIUC}{\affiliation{Department of Astronomy, University of Illinois, 1002 W. Green St., Urbana, IL 61801, USA}}
\newcommand{\Maryland}{\affiliation{Department of Astronomy, University of Maryland, College Park, MD 20742-2421, USA}}

\shorttitle{OTTER}
\shortauthors{Franz \& Friends}

\correspondingauthor{Noah Franz}
\email{nfranz@arizona.edu}

\author[0000-0003-4537-3575]{Noah Franz}
\email{nfranz@arizona.edu}
\altaffiliation{NSF Graduate Research Fellow}
\UA

\author[0000-0002-8297-2473]{Kate D Alexander}
\email{kdalexander@arizona.edu}
\UA

\author[0000-0001-6395-6702]{Sebastian Gomez}
\email{sebastian.gomez@austin.utexas.edu}
\UT

\author[0000-0003-0528-202X]{Collin T Christy}
\email{collinchristy@arizona.edu}
\UA

\author[0000-0003-1792-2338]{Tanmoy Laskar}
\email{tanmoy.laskar@utah.edu}
\Utah

\author[0000-0002-3859-8074]{Sjoert van Velzen}
\email{sjoert@strw.leidenuniv.nl}
\Leiden

\author[0000-0003-1714-7415]{Nicholas Earl}
\email{nmearl2@illinois.edu}
\UIUC

\author[0000-0003-3703-5154]{Suvi Gezari}
\email{sgezari@stsci.edu}
\Maryland

\author[0000-0003-2495-8670]{Mitchell Karmen}
\altaffiliation{NSF Graduate Research Fellow}
\email{mkarmen1@jhu.edu}
\JHU

\author[0000-0003-4768-7586]{Raffaella Margutti}
\email{rmargutti@berkeley.edu}
\UCB
\affiliation{Department of Physics, University of California, 366 Physics North MC 7300,
Berkeley, CA 94720, USA}
\affiliation{Berkeley Center for Multi-messenger Research on Astrophysical Transients and Outreach (Multi-RAPTOR), University of California, Berkeley, CA 94720-3411, USA}

\author[0000-0002-0744-0047]{Jeniveve Pearson}
\email{jenivevepearson@arizona.edu}
\UA

\author[0000-0002-5814-4061]{V. Ashley Villar}
\email{ashleyvillar@cfa.harvard.edu}
\CfA
\affiliation{The NSF AI Institute for Artificial Intelligence and Fundamental Interactions, USA}

\author[0000-0001-6047-8469]{Ann I Zabludoff}
\email{aiz@arizona.edu}
\UA

\begin{abstract}
  Multiwavelength analyses of astrophysical transients are essential for understanding the physics of these events. To make such analyses more efficient and effective, we present the Open mulTiwavelength Transient Event Repository (OTTER), a publicly available catalog of published transient event metadata and photometry. Unlike previous efforts, our data schema is optimized for the storage of multiwavelength photometric datasets spanning the entire electromagnetic spectrum from multiple published sources. Open source software, including an application programming interface (API) and web application, are available for viewing, accessing, and analyzing the dataset. For the initial release of OTTER, we present the largest ever photometric archive of tidal disruption event (TDE) candidates, including $\gtrsim \added{118,000}$ observations of \ntdes TDE candidates spanning from radio to X-ray wavelengths. We demonstrate the power of this infrastructure through four example analyses of the TDE population. We plan to maintain this dataset as more TDE candidates are proposed in the future and encourage other users to contribute by uploading newly published data via our web application. The infrastructure was built with the goal of archiving additional transient data (supernovae, gamma-ray bursts, fast blue optical transients, fast radio bursts, etc.) in the future. The web application is available at \url{https://otter.idies.jhu.edu} and the API documentation is available at \url{https://astro-otter.readthedocs.io}.   
\end{abstract}


\section{Introduction} \label{sec:intro}

\begin{figure*}
    \centering
    \includegraphics[width=0.9\linewidth]{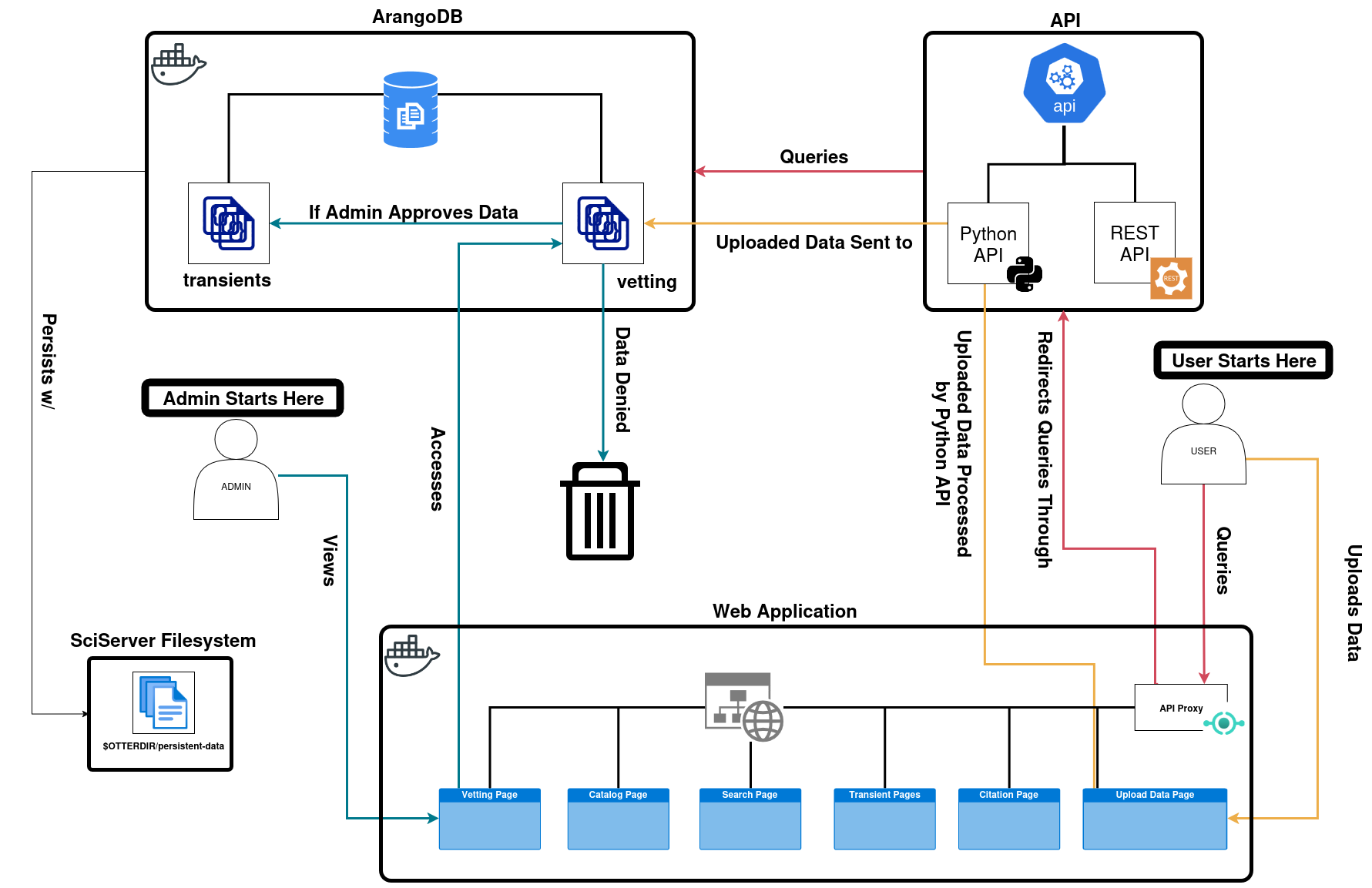}
    \caption{Diagram of the OTTER infrastructure. The infrastructure is made up of a front end web application (bottom right), application programming interface (API; top right), and backend database (left). The different colored arrows show different workflows, two of which are for users (red and yellow) and one for administrators (blue). All of the infrastructure is containerized using docker and deployable on a kubernetes cluster.}
    \label{fig:otter-diagram}
\end{figure*}

The study of astronomical transients presents a unique view of the Universe through an indispensable high-energy laboratory that is impossible to recreate on Earth. Examples of transient events include 1) core collapse supernovae (CCSNe), when a massive star ends its life in a violent explosion \citep[radiated isotropic energy, $E_{\rm iso,\,rad}\sim10^{51}$ erg;][]{1966ApJ...143..626C,1995ApJS..101..181W,smartt_progenitors_2009}; 2) thermonuclear supernovae (e.g., Type Ia SNe), when a degenerate white dwarf star explodes \citep[$E_{\rm iso,\,rad} \sim 10^{51}$ erg;][]{1966ApJ...143..626C, maoz_observational_2014}; 3) gamma-ray bursts \citep[GRBs;][]{kouveliotou_identification_1993,norris_frequency_1984}, hypothesized to occur when a massive star collapses and produces a jet \citep[]{2003ApJ...591..288H} and/or when two neutron stars merge \citep[][$E_{\rm iso,\,rad} \sim 10^{54}$ erg]{eichler_nucleosynthesis_1989,narayan_gamma-ray_1992,abbott_gravitational_2017, savchenko_integral_2017, goldstein_ordinary_2017, fong_electromagnetic_2017}; and 4) tidal disruption events (TDEs; $E_{\rm iso,\,rad}\sim 10^{50-54}$ erg), when a star is torn apart by a supermassive black hole \citep[SMBH;][]{hills_possible_1975, rees_tidal_1988}.

Most past transient catalogs were focused on a specific wavelength regime (e.g., only ultraviolet/optical/infrared, or radio, or X-rays) and resulted in an immense amount of information about the populations \citep[e.g., ][]{2010AJ....139..120F, 2011MNRAS.412.1441L, 2012MNRAS.425.1789S, 2018ApJ...859..101S, van_velzen_seventeen_2021, hammerstein_final_2023, yao_tidal_2023, gomez_type_2024, perley_zwicky_2020, 2023MNRAS.520.4356N, somalwar_vlass_2023, somalwar_vlass_2023-1, dykaar_untargeted_2024, cendes_ubiquitous_2024, 2024A&A...692A..95A}. However, transient events inherently emit across the electromagnetic spectrum, making {\it multiwavelength analyses of time-series datasets} essential to accelerate our astrophysical understanding of these high energy explosions \citep[e.g., ][and many more]{costa_discovery_1997, olivares_e_multiwavelength_2015, gezari_x-ray_2017, pasham_discovery_2018, eftekhari_radio_2018, eftekhari_late-time_2021, laskar_first_2022, laskar_radio_2023, margutti_luminous_2023, gomez_type_2024, christy_peculiar_2024, hajela_eight_2024, guolo_systematic_2024, masterson_new_2024, alexander_multi-wavelength_2025}. As a result, select past catalogs also included limited multiwavelength data \citep[e.g.,][]{auchettl_new_2017, fong_short_2022, nugent_short_2022, langis_repeating_2025}. However, the multiwavelength support in these catalogs remained limited because curating large multiwavelength catalogs of transient events can be particularly laborious, taking {\it years} to gather the data and clean it into a coherent dataset. To bridge this gap, we created the Open mulTiwavelength Transient Event Repository (OTTER) --- an API (application programming interface) first catalog and web-application {\it designed} for archiving large multiwavelength datasets.

OTTER is a successor to the original Open Astronomy Catalogs \citep[OAC; i.e. the Open Supernovae Catalog and Open TDE Catalog;][]{guillochon_open_2017, auchettl_new_2017}, which have not been maintained since $\sim 2018$. Building off the OACs, we designed our infrastructure to both inherently store multiwavelength data and scale efficiently with large all sky surveys. In particular, we have built a flexible dataset structure (i.e., a ``data schema'') to explicitly store information about the nuances related to X-ray and radio data in the catalog (e.g., the fact that X-ray fluxes are model dependent, see \S\ref{sec:xray-data-cleaning}). Additionally, we incorporate the ability for each transient to have multiple classifications with descriptive flags to allow for ease of sample selection (e.g., a flag that tells the user if the event is spectroscopically classified, or another flag that tells the user if the classification is debated). This is important to account for transients that change their classifications because of new observational evidence \citep[e.g., AT\,2017bcc, which was initially classified as a TDE but later reclassified as an ambiguous nuclear transient;][]{2017TNSCR.221....1B, ridley_time-varying_2024} or transients that appear to transition between classifications over time \citep[e.g., SN PTF11iqb, which observationally evolved from a Type IIn to Type IIP/L at late times; ][]{smith_ptf11iqb_2015}.   

This paper presents an initial release of the OTTER data storage and analysis tools for published transient metadata and photometric datasets. This release includes both a web application and API for data access. We would like to emphasize that, while we currently focus on transient metadata and photometry, the infrastructure we have developed can also handle other types of transient data (e.g., host galaxy photometry and/or transient spectra\footnote{While the OTTER framework can handle spectroscopy, we defer transient spectroscopic data storage and curation to WISeREP \citep[]{2012PASP..124..668Y}.}) and we expect it to scale efficiently. 

For this first release, we focus on ingesting and analyzing data from TDE candidates. We curate this dataset by directly scraping photometry from pre-existing publications, rather than re-reducing it ourselves. Observations of TDE candidates exist across the electromagnetic spectrum, making them a good starting point for the multiwavelength capabilities of OTTER. As a result, each transient currently in OTTER has, at least at one time, been classified as a TDE or TDE candidate. To illustrate the types of analyses facilitated by OTTER, we also present several worked examples focused on exploring the TDE population and comparing to theory.

To provide context for these examples, we briefly introduce the current theories for the observed emission produced following a TDE. After the star is torn apart by tidal forces, $\sim$half of the material remains bound to the SMBH and forms a thin tidal stream \citep[]{hills_possible_1975, rees_tidal_1988}. This stream of material can then collide with itself in stream-stream collisions, producing shocks and losing energy, and eventually losing enough angular momentum to circularize into an accretion disk around the SMBH \citep[]{guillochon_hydrodynamical_2013,piran_disk_2015,jiang_prompt_2016,bonnerot_formation_2021,lu_self-intersection_2020, dittmann_analytical_2022,huang_bright_2023, steinberg_streamdisk_2024}. The exact circularizaton process and disk formation timescale remain hotly debated  \citep[]{guillochon_hydrodynamical_2013, metzger_bright_2016, roth_x-ray_2016, dai_unified_2018, mockler_weighing_2019, lu_self-intersection_2020, metzger_cooling_2022, bonnerot_formation_2020}.

TDEs produce radiation spanning the full electromagnetic spectrum from radio to X-rays. During disruption and circularization there are numerous potential emission production mechanisms, but the physical origin of much of the emission is still unclear \citep[]{alexander_radio_2020, gezari_tidal_2021}. Both thermal and hard X-ray emission components have been observed in TDEs; thermal X-rays are likely produced from the accretion disk, while hard X-rays are produced by Compton up-scattering of the soft X-rays \citep[]{roth_x-ray_2016, mummery_tidal_2021, guolo_systematic_2024}. Ultraviolet/optical/infrared (UVOIR) emission is frequently (but not always) seen and could be produced by 1) thermal emission produced after the the stream-stream shocks heat the circumnuclear medium \citep[the ``stream-stream model'';][]{piran_disk_2015, shiokawa_general_2015, lu_self-intersection_2020, jiang_prompt_2016, bonnerot_formation_2021, dittmann_analytical_2022, steinberg_streamdisk_2024}, or 2) reprocessing of X-rays into the UVOIR wavelengths by intervening material shielding the accretion disk \citep[the ``reprocessing model'';][]{loeb_optical_1997, guillochon_hydrodynamical_2013, metzger_bright_2016, roth_x-ray_2016, dai_unified_2018, mockler_weighing_2019, lu_self-intersection_2020, metzger_cooling_2022}. Finally, radio synchrotron emission is seen in $\sim50\%$ of TDEs and could be produced by outflows and/or jets (that could be launched at various points throughout the process) or the unbound stellar debris interacting with the ambient medium \citep[]{giannios_radio_2011, krolik_2016, piro_late-time_2025}. It is possible to disentangle these models only when observational analyses include the full spectrum of multiwavelength information on the TDE, further necessitating multiwavelength catalogs like OTTER \citep[e.g.,][]{pasham_discovery_2018, hajela_eight_2024, alexander_multi-wavelength_2025}.

This paper is organized as follows. In \S\ref{sec:data} we describe the dataset schema, the data cleaning process, the database backend, and the classification flagging. In \S\ref{sec:contrib}, we outline the website and the API, and describe how to contribute new datasets to the catalog as they get published. In \S\ref{sec:current-dataset}, we summarize the current state of the dataset. In \S\ref{sec:sample}, we outline a few applications of the OTTER dataset and API to the population of TDE candidates currently stored in OTTER. We conclude and discuss future directions in \S\ref{sec:conclusion}. Appendix \ref{app:meta} lists the metadata and references for all of the transients currently in OTTER. Appendix \ref{app:schema} extends \S\ref{sec:data} with a detailed description of the dataset schema. Appendix \ref{app:non-cumulative-hist} includes additional plots of the current OTTER dataset.


\section{Methodology \& Data Organization}\label{sec:data}

The primary objectives of OTTER are:
\begin{enumerate}
    \item A complete dataset of transient events that have at least one associated {\it refereed publication}, including their photometry and meta data (names, coordinates, distances, etc.). In this work we start with TDE candidate data but with the plan to expand OTTER to include other transient events in the future.
    \item Publicly available datasets that are easy to access via an API and a web application.
    \item An easy-to-use upload form for anyone to contribute data, provided the data is published.
    \item Multiple classification flags for each included transient, including an overall confidence flag and intuitive data filtering options, to make it easy for users to curate a sample optimal for their science case from the broader OTTER catalog.
    \item The ability to store data from any wavelength.
    \item Appropriate and correct citations, and making it easy for others to appropriately cite the original sources when using data from OTTER.
    \item Scalability and longevity: We hope to make this a long-lasting catalog/software that will scale appropriately as more transients are discovered. In the future, this will likely include transients with an ambiguous nature and probabilistic classifications.
\end{enumerate}

\noindent To accomplish these goals, there are four parts to the OTTER data and infrastructure organization (\figref{otter-diagram}):
\begin{enumerate}
    \item {\bf Cleaned Dataset:} Similar to the OAC, and for longevity, we store all of our data in a human readable JSON format. The raw JSON files are currently available on GitHub\footnote{\url{https://github.com/astro-otter/otter-backups}} and will be made available on Zenodo as the dataset size grows. These JSON files are described in \secref{schema}. To curate a clean dataset of TDE candidate photometry, we pull raw datasets from papers, public catalogs, and data uploaded by external users to the OTTER website and run them through our data cleaning pipeline, described in \secref{pipeline}.
    \item {\bf Backend Database:} We use a database server to make the files programmatically accessible for both the API and the website. This is described in \secref{db}. This is hosted by Johns Hopkins' SciServer \added{\citep[]{taghizadeh-popp_sciserver_2020}}\footnote{\url{https://www.sciserver.org/}}.
    \item {\bf Website:} The OTTER website provides an easy view of the data available in OTTER. This is described in \secref{web}. This is also hosted by Johns Hopkins' SciServer \added{\citep[]{taghizadeh-popp_sciserver_2020}}.
    \item {\bf API:} There is a Python API, built on the more general RESTful API\footnote{\url{https://docs.arangodb.com/stable/develop/http-api/}}, that allow user access to the cleaned dataset. This is described in \secref{api} and in the API documentation\footnote{\url{https://astro-otter.readthedocs.io/}}. The API documentation also includes multiple Python Jupyter notebooks with examples of using the API to query the database\footnote{\url{https://astro-otter.readthedocs.io/en/latest/examples/welcome.html}}.
\end{enumerate}

\subsection{Data Schema}\label{sec:schema}
The JSON format accommodates a wide variety of data types and values by storing nested key-value pairs, while also allowing anyone to open and read the data with any text editor. All of the data associated with each transient is stored in a single, standardized, JSON document. The JSON schema flexibility allows us to add documented sub-keywords depending on the type of data stored. The JSON schema is outlined in detail in Appendix \ref{app:schema} and in the API documentation, but we provide a summary of the schema organization here. 

Our JSON file schema is broken into ``properties,'' ``subproperties,'' and ``keywords.'' The ``properties'' are the primary keys of our standardized JSON file schema, and are \texttt{name}, \texttt{coordinate}, \texttt{distance}, \texttt{date\_reference}, \texttt{classification}, \texttt{host}, \texttt{photometry}. If the property is a JSON dictionary, ``subproperties''  are the sub-keys used within properties. For example, the \texttt{alias} is a subproperty of the \texttt{name} property which stores multiple names associated with a single transient. ``keywords'' are the keys of the dictionaries stored under ``subproperties'' (when the property is a dictionary that stores subproperties) or ``properties'' (when the property itself stores a list of dictionaries). 

The only primary keys that are required in every JSON document are \texttt{name} and \texttt{coordinate}. All other properties are optional and included if the information is available. Importantly, every value of these primary keywords has a list of NASA Astrophysics Data System (ADS\footnote{\url{https://ui.adsabs.harvard.edu/}}) bibcodes associated with it so that each value can be appropriately cited with the original source. 

If the \texttt{photometry} property is provided, we require that each observation reported has keywords corresponding to the flux value and units; observation date value and units; and filter name, wavelength/frequency, and wavelength/frequency units. In addition, if any corrections were applied to the photometry (host subtractions, k-correction, dust corrections, etc.), we require that the relevant corrections are provided to ensure reproducibility of the photometry point.

We currently only store published photometry in OTTER with the expectation that spectra will be uploaded to WISeREP \citep{2012PASP..124..668Y}\footnote{\url{https://www.wiserep.org/}} when published. We also store only limited host galaxy information, including a name and coordinates. This is to avoid storing duplicate information with other galaxy archives. However, in our Python API we provide easy methods for querying public catalogs for additional host photometry and archival spectra.

\subsection{Data Backend} \label{sec:db}
The data is stored in an ArangoDB document database\footnote{\url{https://arangodb.com/}}. Unlike most other publicly accessible astronomical databases \citep[e.g.,][]{guillochon_open_2017, lindstrom_tom_2022, hosseinzadeh_saguaro_2024, jones_blast_2024}, we chose to use a document database over a more classic relational database management system (like MySQL, PostgreSQL, etc.)\footnote{Even though they are relatively unheard of in astronomy, document databases are surprisingly popular in everyday use. For example, ArangoDB is used by companies like Cisco, Deloitte, Mercedes-Benz, and more! And, other, closed-source, document databases are used by companies like Netflix, Uber, and Airbnb.}. We made this choice because document databases provide a more flexible storage schema where we are able to have unlimited optional keywords that are required to have a value. Such a flexible database storage schema is essential for datasets like ours where the data keyword changes depending on the types of values stored (e.g., the storage of radio photometry compared to X-ray photometry) and the public availability of optional keywords (e.g., the telescope used to take the data). Furthermore, the document database structure provides an intuitive way to store and query default and non-default values for a quantity from different sources. For example, the document database schema can store multiple different redshifts that were derived by different teams for a single transient event. 

In terms of scalable data storage, ArangoDB will automatically shard\footnote{This means that the database will automatically distribute the dataset across multiple storage nodes.} the database documents across multiple storage nodes making it one of the most scalable options for our dataset. Furthermore, the ArangoDB query language (AQL) provides a very fast way to administer ``typical" queries of our dataset, even as the dataset grows orders of magnitude larger. Finally, the document database structure is essentially identical to the JSON format, making uploading the cleaned dataset extremely efficient. For all of these reasons, a document database is essential to provide the scalable infrastructure necessary as the OTTER dataset grows larger.  

Unfortunately, this decision comes at a minor reduction in the ease of access to the dataset. As mentioned above, the document database structure requires the use of the more flexible AQL rather than the more commonly known SQL (i.e., structured query language). This means that people who wish to query the database directly must learn a new query language. We try to address this problem by providing a comprehensive search form on our website and a Python API. Nevertheless, if someone would like to query the database directly using AQL, the language is open source and has publicly-available documentation\footnote{https://docs.arangodb.com/stable/aql/}. Therefore, we decided the benefits of using a document database significantly outweigh the potential downsides.

\subsection{Data Sources \& Cleaning Pipeline} \label{sec:pipeline}

To create the initial OTTER TDE candidate catalog, we compile data from multiple existing catalogs and individual papers. In addition to cleaning the data into a standard format, each component of each dataset is appropriately cited with an ADS bibcode in the individual JSON files. All of these citations are given in \tabref{refs}. In \secref{otc} through \secref{tns-data} we describe dataset specific nuances for the data cleaning process. 

Note that, for optical photometry without a specified calibration system, we use the filter to make a reasonable assumption about the calibration system. For photometry using SDSS, ATLAS, Swift\footnote{We understand that this may not always be the case for Swift photometry, but we had to choose a ``default'' system when the original source did not specify the magnitude system they calibrated against. This currently only affects 280 ($\lesssim0.35\%$) photometry points stored in OTTER.}, or HST filters we assume they are in the AB system. For photometry using Johnson/Cousins or Wide-field Infrared Survey Explorer \citep[WISE; ][]{WISE} filters, we assume they are calibrated using the Vega system.

Throughout the data cleaning process, we deduplicate identical data (metadata and photometry) by appending the new citation to the previously stored data. This ensures that all appropriate works are cited when the data is used. An intentional artifact of only deduplicating identical datasets is that it does not remove different reductions of the same photometry, allowing the user to choose which flux value they prefer. To prevent users from accidentally using multiple photometry points produced from different reductions of the same data, we provide a default deduplication algorithm as part of the Python API. See \S\ref{sec:api} for a discussion of this algorithm.

\subsubsection{The Open TDE Catalog} \label{sec:otc}

The Open TDE Catalog was, like the Open Supernova Catalog, part of the AstroCats movement to create a standardized data format for public time domain data, including published and unpublished datasets \citep[]{guillochon_open_2017,auchettl_new_2017}. Until $\sim 2022$, this was the most comprehensive catalog of TDE data, with data on $\sim100$ TDE candidates. The data organization in OTTER is influenced by the AstroCats movement, which also used the JSON format for their backend data storage. 

There are some important notes from the conversion of this dataset to the OTTER format. \citet{guillochon_open_2017} were unable to find the photometric corrections that were applied to some of their dataset. As discussed above, this does not align with our requirements for the photometry keywords. But, we still found it necessary to include this dataset for completeness. As a result, for this dataset only, we adopted the standard that if it is unclear whether photometric corrections were applied to the data, we will store ``null" for all of the correction boolean values (see Appendix \ref{app:schema}). We attempt to prevent users from treating this data as corrected by introducing warning messages to both the API and web application whenever relevant. Any of these values can easily be filtered out after obtaining the cleaned photometry from OTTER. 

The Open TDE Catalog also accepted unpublished information and included it in the catalog. This is useful for creating a complete dataset. However, it eliminates a level of vetting done in the peer review process and reduces our confidence in the data quality. Therefore, we decided to only include photometry points from The Open TDE Catalog that have an ADS bibcode. This resulted in a loss of $8.35\%$ of the photometric points from The Open TDE Catalog. 

During this data processing, we found that the Open TDE Catalog had two sets of duplicated entries where the same TDE was stored under two names in two separate files. The first set of duplicated files are for PS18kh and AT\,2018zr and the second set of duplicated files are for J134244 and SDSSJ1342. OTTER handles matching transients and finding duplicated entries by performing coordinate matches to ensure we are not storing data under two separate names. Therefore, these duplicated files are merged in the OTTER data cleaning pipeline.

Similarly, we note that five events from the Open TDE Catalog have not yet been classified in a publication: AT\,2020zeb, ASASSN-20jr, ASASSN-20il, AT\,2021msu, ASASSN-20lk. Additionally, these events have no published photometry. As a result, we remove these from the OTTER sample.

\citet{goldtooth_census_2023} provided essential work to summarize and expand The Open TDE Catalog with new X-ray selected TDEs. This includes the relatively new X-ray selected TDEs discovered by the extended ROentgen Survey with an Imaging Telescope Array (eROSITA) Spektrum-Roentgen-Gamma (SRG) \citep{sazonov_first_2021}, although the X-ray fluxes for these TDEs is unfortunately not publicly available until the SRGE data release. \citet{goldtooth_census_2023} carefully classified each transient from The Open TDE Catalog and gave important host information, both of which we ingest into OTTER. We pull the machine readable file provided by \citet{goldtooth_census_2023}, which is derived from their Table 1, and process the metadata into the OTTER schema. 

\subsubsection{Ultraviolet, Optical, and Infrared Data (UVOIR)}
The so-called ``Curated Optical TDE Catalog" is a continuously updated catalog of optically-selected TDE photometry available at \url{https://github.com/sjoertvv/manyTDE}, with the initial catalog and data curation methods described in \citet[]{mummery_fundamental_2024}. This dataset includes more recent optical/UV/IR photometry for optically-selected TDEs, most of which were discovered after the deprecation of The Open TDE Catalog. Most of this photometry is from the Zwicky Transient Facility \citep[ZTF;][]{ZTF} forced photometry server, WISE
\citep[][]{WISE, NEOWISE}, and the Neil Gehrels Swift Ultraviolet/Optical Telescope (Swift UVOT), with some additional photometry from other optical all sky surveys \citep[]{ASASSN, ATLAS, iPTF, panstarrstde}. The catalog was initially assembled for the analysis performed in \citet{mummery_fundamental_2024}, using methods outlined in \citet{van_velzen_late-time_2019} and \citet{van_velzen_first_2019}, but has grown since then. All of the data is host subtracted and we store the appropriate host subtraction in the OTTER schema. Every time our data cleaning pipeline runs it pulls changes from The Curated Optical TDE Catalog and incorporates the photometry into OTTER. 

The Curated Optical TDE Catalog provides a large majority of the transients in OTTER with a classification (as a TDE), redshift, velocity dispersion, optical peak date, and optical photometry. Note that this dataset has more extensive host photometry and model result parameters than what we currently store in OTTER. Therefore, we leave it to the excited scientist to visit their dataset directly for these values.

\citet{masterson_new_2024} performed an archival search of WISE Infrared (IR) data for flares associated with potential tidal disruption events and found 12 new IR bright TDE candidates. We include their entire photometric dataset, spanning from IR to X-rays, in OTTER as well as specific metadata information like coordinates, redshift, and best-fit disruption time. This dataset was provided through private communication with the authors of \citet{masterson_new_2024}.

We include additional optical photometry not included in the Curated Optical TDE Catalog and \citet{masterson_new_2024}. This includes optical data for Sw J1644+57 \citep{levan_extremely_2011,levan_late_2016}, AT2022cmc \citep[]{andreoni_very_2022}, ASASSN-15oi \citep[]{hajela_eight_2024}, GNTJ145301.70+422127.82 \citep{kostrzewa-rutkowska_gaia_2018}, and AT2017bcc \citep{ridley_time-varying_2024}. We also gathered additional public photometry for AT2021qxv, AT2021ehb, AT2020opy, AT2020nov, and AT2020neh from the Young Supernova Experiment \citep[YSE, ][]{aleo_young_2023}.

\subsubsection{X-ray Data}\label{sec:xray-data-cleaning}
X-ray data is distinct from Radio and UV/Optical/IR (UVOIR) data in that extracting a flux value from the observation requires modeling the observed X-ray spectrum. This means that the derived flux value is model dependent. To ensure that the X-ray fluxes reported in OTTER are reproducible, we require that the model name and best-fit model parameters are stored with the X-ray flux values (with only a few exceptions, see below). In the cases where X-ray spectra are stacked for modeling, we report the same model for all of the extracted flux values.

\citet{guolo_systematic_2024} presented an extensive re-reduction and modeling of X-ray observations of 17 TDEs from Neil Gehrels Swift Observatory, XMM-Newton, and SRG/eROSITA. To extract fluxes from the X-ray spectra, \citet{guolo_systematic_2024} applied either the tdediscspec model or a simPL$\otimes$tdediscspec model \citep{mummery_tidal_2021}. The tdediscspec model describes the thermal X-ray emission produced by a TDE accretion disk. The simPL$\otimes$tdediscspec model is a convolution of a simple comptonization model with the tdediscspec model \citep[]{1999ascl.soft10005A, 2009PASP..121.1279S}. 

We obtained the complete light curves \citep[Figure 2]{guolo_systematic_2024} through private communication and converted all of the data to the OTTER schema. This accounts for any models in which they had to stack multiple XMM-Newton observations to fit. In doing this, we also find that $\sim1/3$ of the models used for extracting the flux are not explicitly stated or cited. As a result, we store null values for those \texttt{xray\_model} subkeyword in the photometry schema. 

Besides the X-ray data from \citet{guolo_systematic_2024}, \citet{masterson_new_2024}, and the Open TDE Catalog, we also include data on ASASSN-15oi from \citet{hajela_eight_2024} and on AT\,2017bcc from \cite{ridley_time-varying_2024}. The X-ray data on both of these events is incorporated into the existing photometry in the database. The models used to extract the fluxes from the X-ray spectra are explicitly provided in both papers and are documented in the \texttt{xray\_model} subkeyword. 

\subsubsection{Radio Data}
A major novelty of this work is the centralization and presentation of all of the publicly available radio data for confirmed TDEs. We scrape numerous single object TDE papers (see Table \ref{tab:refs}), and a few radio population papers  \citep{alexander_radio_2020,somalwar_vlass_2023,dykaar_untargeted_2024,cendes_ubiquitous_2024,anumarlapudi_radio_2024}, for both metadata and photometry.  If the image is host subtracted, we include this in the \texttt{corr\_host} keyword in the photometry, similar to the UVOIR dataset. Any reported uncertainties on the flux calibration related to interstellar scintillation are stored in the \texttt{raw\_err\_detail} and \texttt{value\_err\_detail} keywords using the \texttt{iss} subkeyword.

In general, unlike UVOIR data, radio band names do not have a consistent naming convention. As a result, for all radio data, we attempt to define consistent radio band names for use in OTTER. A detailed description of the frequency ranges used for these bands is given in \autoref{tab:band_names}.

\begin{table*}[]
    \centering
    \caption{Radio Band Name Definitions for OTTER}
    \label{tab:band_names}
    \begin{tabular}{l l c p{3in}}
    \hline
        Band Name & Standard & Frequency Range & Notes \\
         & & [GHz] & \\
        \hline
        HF  & IEEE & (0.003, 0.03] & --- \\ 
        VHF & IEEE &  (0.03, 0.125] & The IEEE sets the official standard as (0.03, 0.3] but we adjust it to match with GMRT standard without overlap.  \\ 
        gmrt.2  & GMRT & (0.125, 0.250] & --- \\ 
        gmrt.3 & GMRT & (0.25, 0.55] & The official GMRT standard is (0.25, 0.5] but we adjust it to match with the IEEE UHF standard \\ 
        UHF & IEEE & (0.55, 1] & ---\\
        L & VLA & (1, 2] & ---\\ 
        S  & VLA & (2, 4] & ---\\
        C  & VLA & (4, 8] & ---\\ 
        X  & VLA & (8, 12] & ---\\
        Ku & VLA & (12, 18]  & ---\\
        K  & VLA & (18, 27] & ---\\
        Ka  & VLA & (27, 40] & ---\\
        alma.1 &ALMA & (40, 50] & The official ALMA standard is (35, 50] but we adjust it to align with the VLA Ka band \\
        alma.3 & ALMA & (84, 116] & For now, we skip alma.2 since it is not yet available for observation \\
        alma.4 & ALMA & (116, 163] & The official range is (125, 163] but we adjust the lower bound to remove the gap between alma.3 and alma.4$^{\hyperref[tabfootnote:a]{a}}$\\
        alma.5 & ALMA & (163, 211] & --- \\
        alma.6 & ALMA & (211, 275] & ---\\
        alma.7 & ALMA & (275, 373] & ---\\
        alma.8 & ALMA & (373, 500] & The official range is (385, 500] but we adjust the lower bound to align with alma.7 $^{\hyperref[tabfootnote:a]{a}}$\\
        alma.9 & ALMA & (500, 720] & The official range is (602, 720] but we adjust the lower bound to align with alma.8$^{\hyperref[tabfootnote:a]{a}}$\\
        alma.10  & ALMA & (787, 950] & ---\\
        \hline
    \end{tabular}
    \begin{minipage}{0.8\textwidth}
        {\footnotesize \label{tabfootnote:a}$^a$These gaps between the original ALMA filter definitions are due to strong water absorption in the atmosphere. As a result, it is unlikely that any data will fall in these ranges. However, we still extend the band definitions here for completeness.}
    \end{minipage}
\end{table*}

\subsubsection{The Transient Name Server}\label{sec:tns-data}

The Transient Name Server (TNS\footnote{\url{https://www.wis-tns.org/}}) is publicly available infrastructure supporting the reporting, naming, and classification of transients. At the end of the data cleaning pipeline, we query the TNS for any other available information on the transients we included in OTTER. We pull all of the relevant metadata information like the coordinates, discovery date, and redshift (if available) from the TNS Daily CSV. For all of this metadata information we cite the NASA ADS bibcode associated with the TNS Astronote where the data originated.    

\subsection{Classification Confidence Flagging}\label{sec:class-conf}
Accurately and efficiently classifying transients, particularly rare transients like TDEs, remains a complex problem and a matter of intense discussion. For example, in the case of TDEs, while the gold standard in the field is often taken to be the existence of a high signal-to-noise optical spectrum with characteristic emission lines (e.g., \citealt{arcavi_continuum_2014}), this excludes TDE candidates that are discovered at other wavelengths (e.g., X-ray, radio) and do not show any optical emission --- even though many of these objects are also widely accepted as TDEs \citep[e.g.,][]{auchettl_new_2017,anderson_caltechnrao_2020}. To help with this ambiguity and facilitate the curation of ``reliable" population sub-samples for different science cases, we provide numerous flags for the classifications associated with any given transient. The classification flags described in this subsection are applicable to most transients that will eventually be included in OTTER, including both TDEs and SNe. However, with the addition of some transient events (e.g., GRBs) in the future, other, more relevant, classification flags may become necessary. These flags also prepare us for a future where many transients will not be optically spectroscopically classified, in which case the classification may remain ambiguous indefinitely.


Since a transient can have multiple classifications from different citations, we flag all classifications associated with a single transient. Each of these classifications has a numerical ``confidence'' flag defined in Table \ref{tab:conf-flag} (where $C$ indicates the confidence flag). In general, a larger $C$ typically means that more astronomers agree on this classification, although this may not always be the case. Transient classifications derived from photometry-based machine learning algorithms will always have $C=1$, but for these we also provide a \texttt{ml\_score} keyword that can be used to assess reliability. 

\begin{table}[!h]
    \centering
    \caption{Classification Confidence Flag Definitions}
    \begin{tabular}{l p{6.5cm}}
         Flag & Definition \\
         \hline
         $C = 0$ & This classification is unverified. This includes sources that have no publicly available photometry. In the specific case of TDEs, this also includes sources where they are likely AGN but at one point someone speculated that they may be a TDE. \\ 
        $C = 1$ & This classification is from a refereed publication with a reasonable degree of confidence but only relies on photometry to make the classification. This, therefore, includes the transients classified by photometry-based ML algorithms and allows a single transient to have multiple ML classifications. In the case of TDEs, this includes the more historical X-ray selected events that do not have spectroscopic classifications during the TDE flare. \\
        $C = 2$ & This classification relies on a published optical spectrum in a non-referred publication (e.g., classified on TNS). \\
        $C \geq 3$ & This classification is based on an optical spectrum published in a referred publication. Any spectroscopic classification will have $C\geq3$, but this then has sub-flagging with a decimal place system to handle the spectroscopic quality classifications used in the \texttt{spec\_classed} meta-level flag: gold is given a flag of 3.3, silver a flag of 3.2, and bronze a flag of 3.1. The \texttt{spec\_classed} meta-level flag is derived based on the presence of a $C \geq 3$ classification. \\
    \end{tabular}
    \label{tab:conf-flag}
\end{table}

\begin{figure*}
    \centering
    \includegraphics[width=0.9\linewidth]{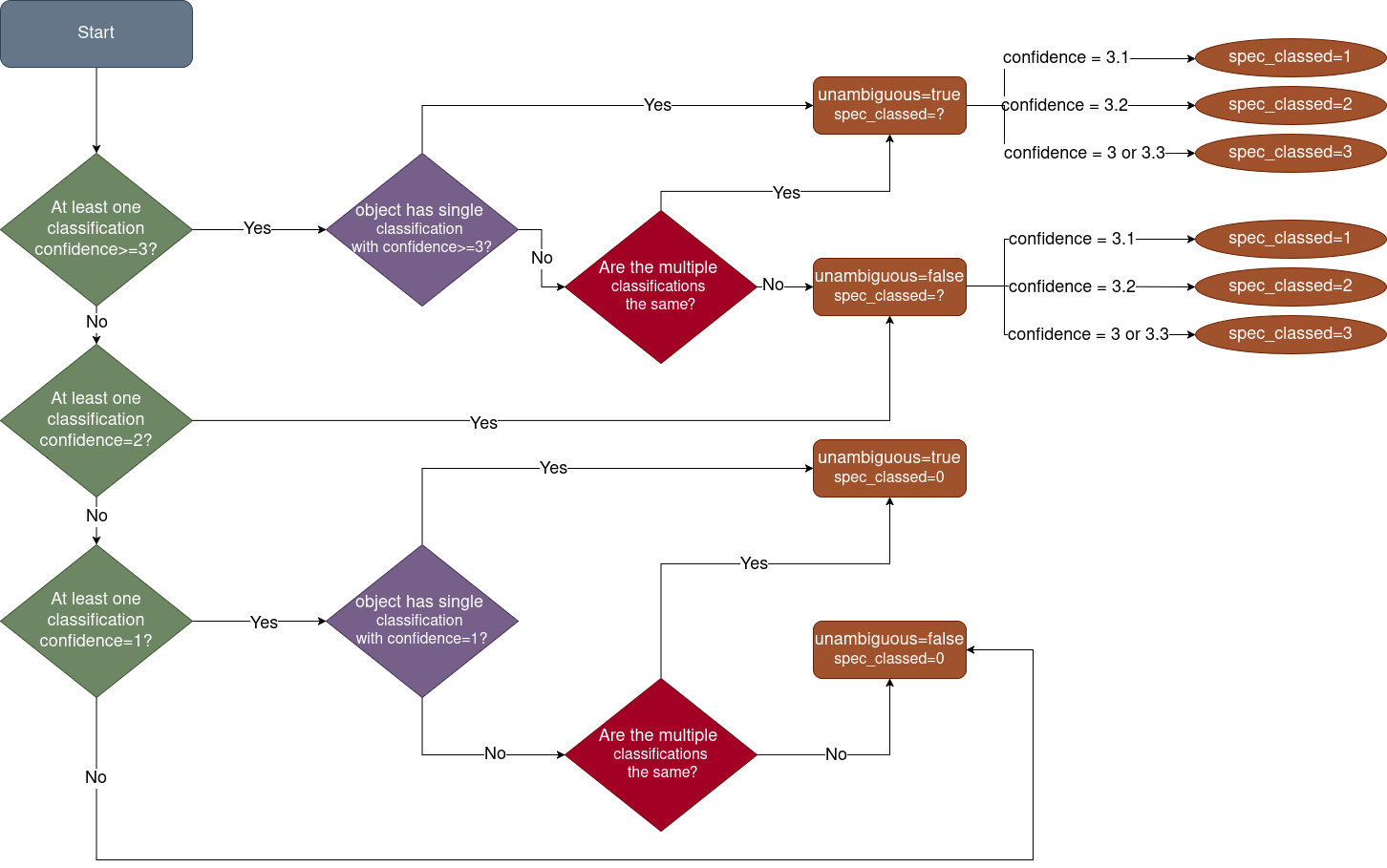}
    \caption{A decision tree showing our process for flagging classifications in OTTER. The classification confidence is a keyword for each individual classification associated with each transient and is then used to apply higher level flags to the classification. The flags are defined in the text and in Table \ref{tab:conf-flag}.}
    \label{fig:flagging-decision-tree}
\end{figure*}

Based on this flagging schema, we can outline the automated algorithm for deriving meta-level flags describing how each transient was classified. \autoref{fig:flagging-decision-tree} is a flow chart that shows this algorithm. Currently, we store three such meta-level flags. The first is simply the discovery method of the transient (\texttt{discovery\_method}). The \texttt{discovery\_method} flag is currently either ``xray'', ``uv'', ``optical'', ``ir'', or ``radio'', but it could be extended to also include gamma-rays, neutrinos, or gravitational waves as future discovery methods. This flag can be particularly useful for selecting TDE samples as the discovery wavelength regime is tied to some astronomers' confidence that the event is a TDE rather than, for example, a SNe.

Many transients are classified based on a common set of optical spectroscopic properties. For example, TDEs are typically classified based on a persistent blue continuum and broad Hydrogen and/or Helium lines \citep{arcavi_continuum_2014, charalampopoulos_detailed_2022}. As a result, the second flag informs the user if the event has been optically spectroscopically classified (\texttt{spec\_classed}), which ranges from 0-3. We use a range since it appears commonplace in the literature to further distinguish classification spectra into bronze (1), silver (2), and gold (3) classes depending on their quality. If the paper does not distinguish their classifications into confidence levels (i.e. bronze, silver, and gold), we assume that the classification is ``gold'' and set \texttt{spec\_classed} to 3. If the transient does not have an optical spectroscopic classification, \texttt{spec\_classed} is set to 0.

The third meta-level flag informs the user if the transient event has any competing classifications by peer-reviewed published sources (\texttt{unambiguous}). The \texttt{unambiguous} flag is set to true if all peer-reviewed sources on the transient agree on a classification. Alternatively, if at least one peer-reviewed source disagrees, the \texttt{unambiguous} flag is set to false. If the source does not have any peer-reviewed, published classifications, the \texttt{unambiguous} flag is set to false.  

\begin{figure}
    \centering
    \includegraphics[width=\linewidth]{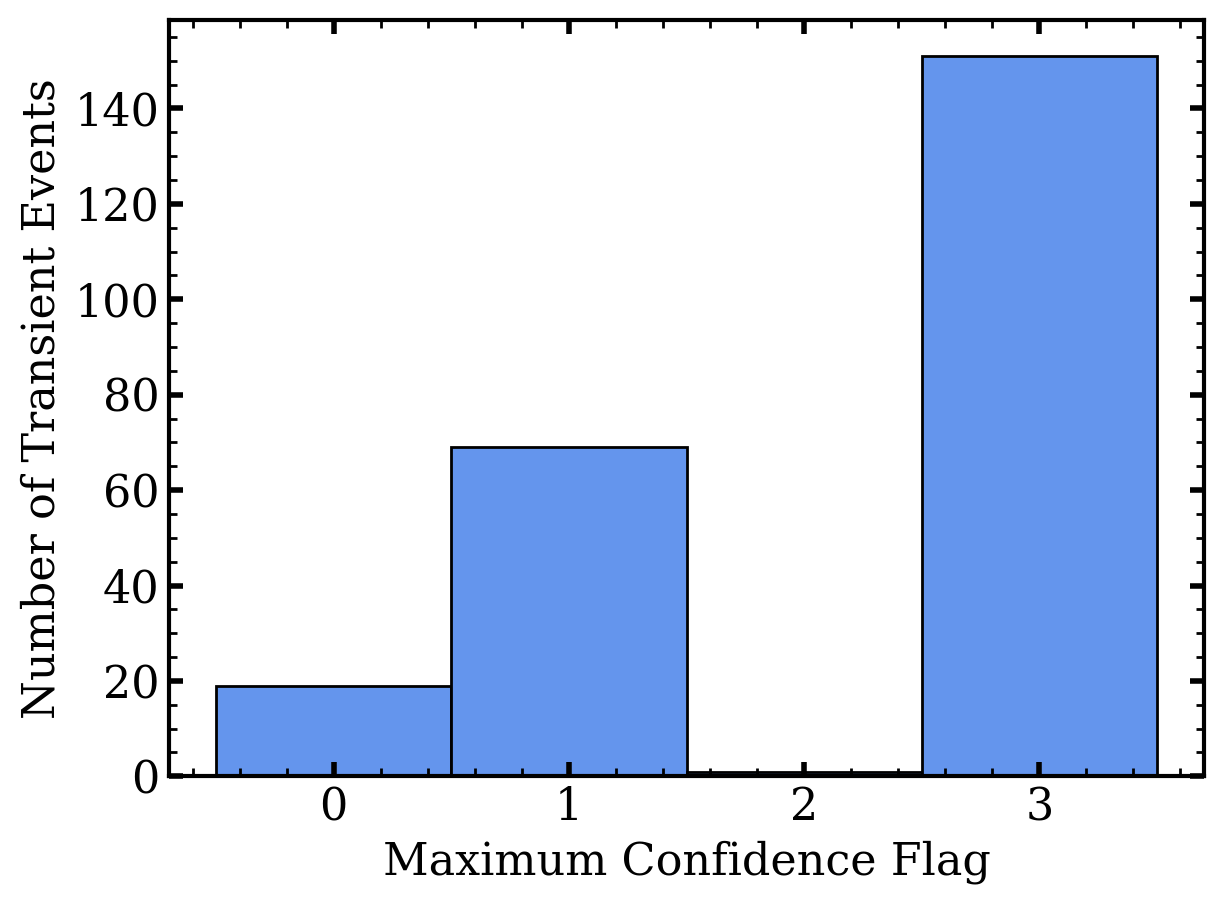}
    \caption{Distribution of the maximum confidence flag we assign to a classification in each transient. Most transient events in OTTER are classified with at least one optical spectrum.}
    \label{fig:max-conf-flag-hist}
\end{figure}

\autoref{fig:max-conf-flag-hist} shows the distribution of the maximum confidence flags associated with each transient event (i.e., TDE or TDE candidate) stored in OTTER. This distribution demonstrates that most ($\sim 150$) the events in OTTER are classified from optical spectroscopy and about a third as many are only photometrically classified. The most pure sample of TDEs in OTTER probably has the flags \texttt{unambiguous} $={\rm true}$ and \texttt{spec\_classed} $\geq1$. However, selecting on these flags does bias the sample to optically-selected TDE candidates, even though there are other highly probable TDE candidates in the catalog. As a result, we leave the exact sample selection to the discretion of the individual user. 

\section{Data Availability \& Contributing Data} \label{sec:contrib}
As described in \S\ref{sec:data}, two of the primary goals of OTTER are 1) allow for the data to be easily accessed and 2) make it easy to contribute data after it is published, with the hope that others will make this contribution part of their typical publishing workflow. As such, we developed two ways to interface with the OTTER dataset, neither of which require any account registration or login: 1) an open access website for easily viewing and uploading datasets and 2) an API for programmatic access to the data. In the following subsections, we describe this infrastructure. 

\begin{figure*}
    \centering
    \includegraphics[width=0.9\linewidth]{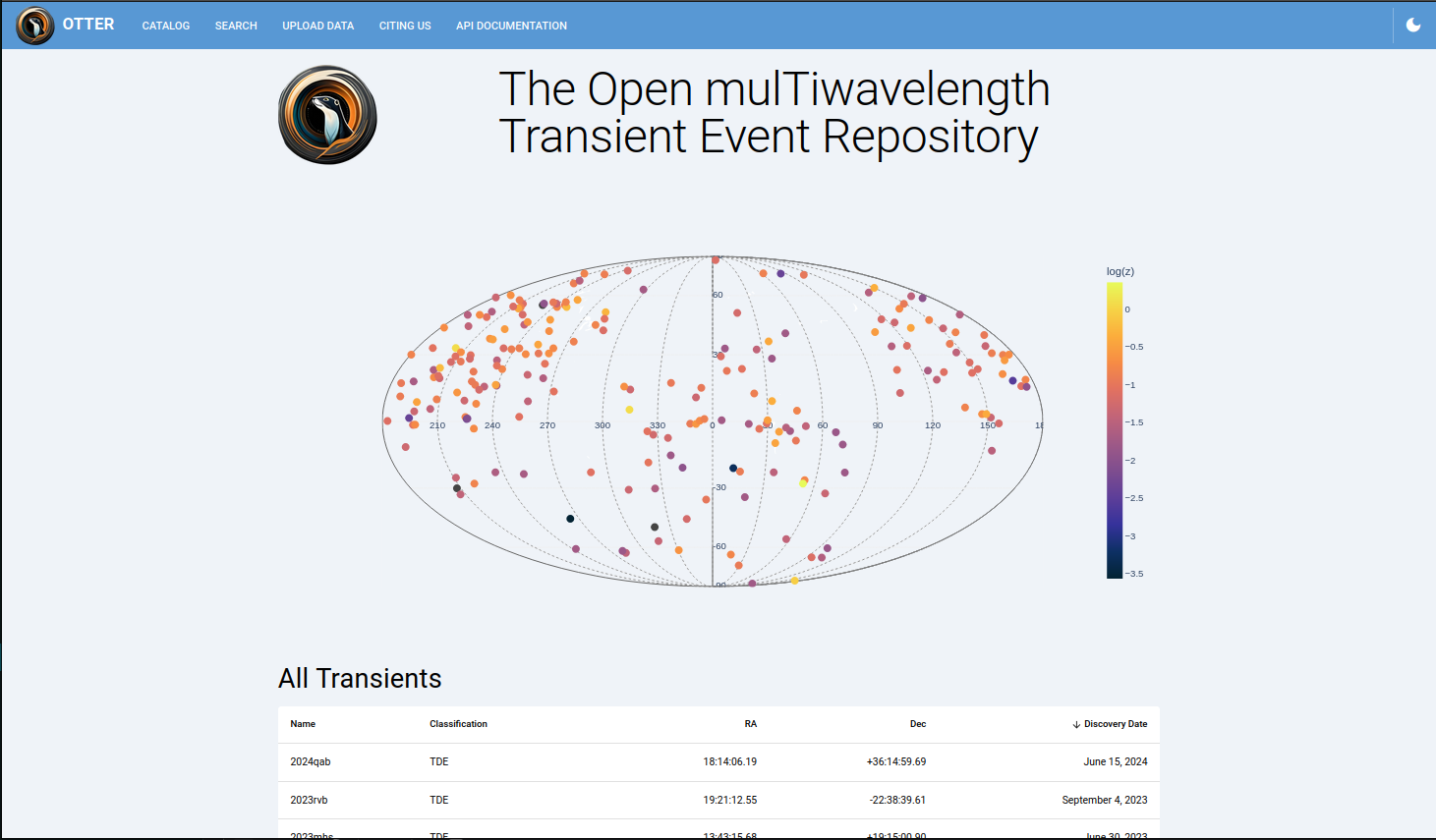}
    \caption{The OTTER home page with the sky map and table of objects. The objects in OTTER are spread relatively uniformly across the sky. There is a slight bias towards the nothern hemisphere due to the high TDE discovery rate of the Zwicky Transient Facility.}
    \label{fig:home-page}
\end{figure*}

\subsection{OTTER Website} \label{sec:web}
The OTTER website takes influence from both the original OACs and the Searches After Gravitational Waves Using ARizona Observatories (SAGUARO) Target and Observation Management (TOM) application \citep[]{guillochon_open_2017, lindstrom_tom_2022, hosseinzadeh_saguaro_2024}. It includes 1) a catalog home page, 2) a search page, 3) individual event pages, 4) an upload form, 5) a citation page, and 6) a vetting page. Every page is publicly accessible without any required login with the exception of the vetting page, which only specific, trusted users can access. This makes the OTTER dataset easy to access while also allowing each uploaded dataset to be carefully vetted by known astronomers.

The catalog home page (\autoref{fig:home-page}) includes a table with metadata and a skymap of the objects in the catalog. The OTTER search page includes two forms, one that allows users to query the catalog based on the transient metadata and a second that allows a user to pass raw AQL queries to the ArangoDB backend\footnote{\url{https://docs.arangodb.com/stable/aql/}}. The query form has a range of search options, including by name (or partial name), coordinate cone search, a redshift range, a classification type, and/or if photometry is available. The results of any of these searches produces a table similar to the home page. The results of the search are also exportable to a zip file of plain text JSON files for each transient so that users can easily access the full datasets they are interested in.

\begin{figure*}
    \centering
    \includegraphics[width=0.7\linewidth]{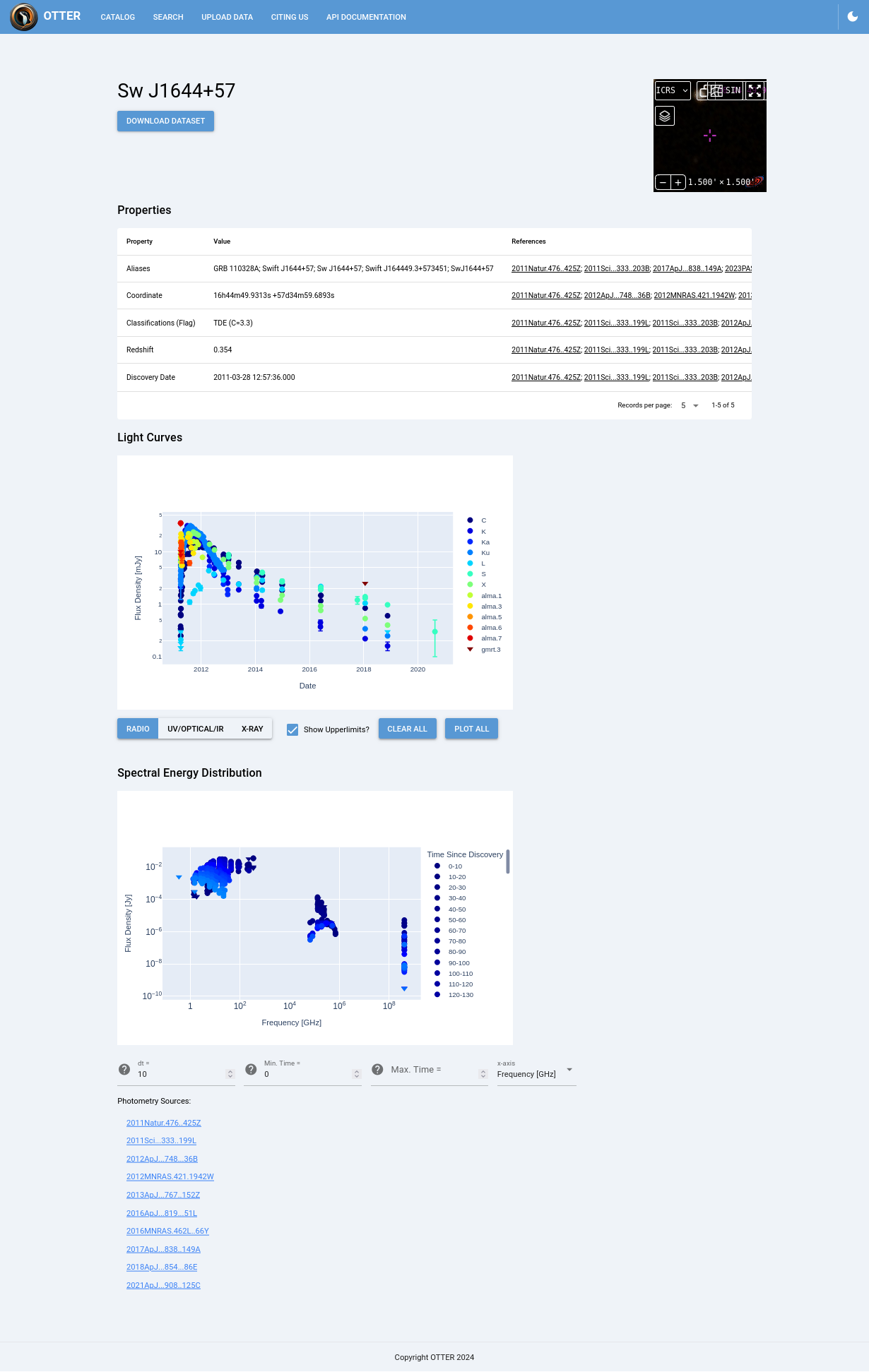}
    \caption{Sample transient page for the TDE Sw J1644+57 \citep[]{2011Natur.476..425Z,
2011Sci...333..199L,
2012ApJ...748...36B,
2012MNRAS.421.1942W,
2013ApJ...767..152Z,
2016ApJ...819...51L,
2016MNRAS.462L..66Y,
2017ApJ...838..149A,
2018ApJ...854...86E,
2021ApJ...908..125C}. The quick look image is from ALADIN \citep[]{2000A&AS..143...33B, 2014ASPC..485..277B, 2022ASPC..532....7B}. These individual transient pages provide basic metadata information on the transient and show the available photometry stored in OTTER. The photometry is displayed with both a light curve (top) split by wavelength regimes and a spectral energy distribution (SED) with tunable hyperparameters. The SED hyperparameters include a $\Delta t$ (used for binning and colorizing the data), a maximum and minimum time to plot, and an X-axis version (Frequency in GHz or Wavelength in nm).}
    \label{fig:transient-page}
\end{figure*}

Each individual transient page includes the names assigned to the transient, a table of its basic properties, and plots of its light curve and spectral energy distribution. A sample of one of these pages is shown in \autoref{fig:transient-page}. The properties table lists what we have deemed to be the ``default'' values for the various transient properties that are available. For most properties, if multiple values are reported in the literature, then the default value is determined by the value that has the most citations. There are two exceptions: 1) the default name of the transient, for which we select the International Astronomical Union convention (i.e. TNS or TNS-like) names over other transient aliases used in the literature (which are also listed in the table) and 2) the default classification, which is based on the highest confidence flag\footnote{see \S\ref{sec:class-conf} for a discussion of the confidence flagging and the biases that this default classification algorithm may introduce.}. The light curve and spectral energy distribution plots are shown if photometry is publicly available. 

\subsubsection{Uploading Data via the OTTER Website}

The data upload page has two available forms. The more simple form is the single object upload form, which allows astronomers to upload their data by providing the metadata and photometry for a single transient. The second form is a bulk upload form where we allow uploading all of the metadata for multiple objects in one file and all of the photometry in a second file. These upload forms attempt to minimize the reformatting necessary for existing datasets while balancing the need for a standard format to add the data to OTTER. 

To help maintain (in the case of TDE candidates) and build (in the case of other transients) a complete dataset in OTTER, we encourage the community to upload their data to OTTER when it is accepted for publication as a standard part of their workflow. The only data we require users to provide is their name, their email, the transient name, the transient coordinates, and the bibcode we should reference for that data. Other optional metadata fields include the redshift, luminosity distance, comoving distance, discovery date, and classification. For each piece of metadata the user is required to provide a citable bibcode.

\begin{figure*}
    \centering
    \includegraphics[width=0.9\linewidth]{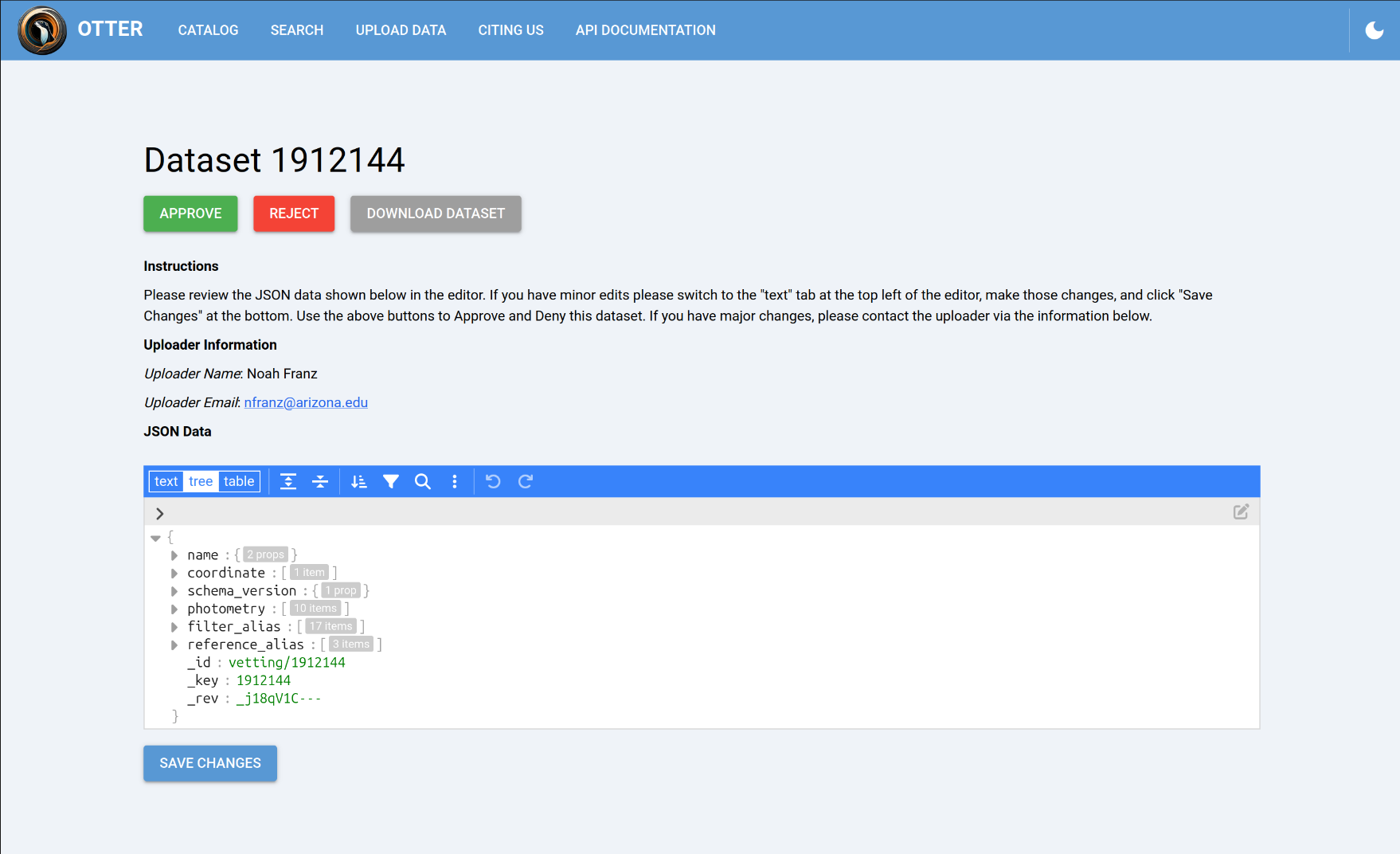}
    \caption{A sample vetting page for a transient dataset upload. This page is the only private page on OTTER and is meant for administrators to view and approve/deny datasets that were uploaded. The data shown in this figure is a sample dataset for a fake TDE.}
    \label{fig:vetting}
\end{figure*}

After a user uploads their data, some automated vetting is performed to ensure that it is valid. The dataset is then made visible to trusted astronomers (i.e. ``vetters'') in a way that allows them to analyze and approve (or reject) the uploaded data. If the data is approved, it is merged (including deduplication of identical datasets) and uploaded to the primary database. 
An example of these vetting pages is shown in \autoref{fig:vetting}.

\subsection{OTTER API} \label{sec:api}
The OTTER API provides programmatic access to the dataset. The Python OTTER API is a pip installable package that gives easy access to astronomers who hope to use this dataset for their research. The full documentation of the OTTER Python API, including examples with common use cases, is at \url{https://astro-otter.readthedocs.io} and is also linked on the OTTER website. We summarize the functionality built into this Python API here.

The OTTER Python API includes codes to query the database and interface with the OTTER JSON documents. This makes it easy to access metadata and photometry on each of the transients. The photometry will be converted to the requested units using astropy affiliated package \texttt{synphot}. The Python API also makes it easy to query other astronomical surveys and databases including Simbad \citep[]{2000A&AS..143....9W}, ATLAS \citep[]{ATLAS}, ZTF \citep[]{ZTF}, iPTF \citep[]{iPTF}, ASAS-SN \citep[]{ASASSN,2017PASP..129j4502K,2023arXiv230403791H}, Vizier \citep[]{10.26093/cds/vizier, vizier2000}, WISE \citep[]{WISE,NEOWISE,NEOWISE_Reactivation,2020MNRAS.493.2271H}, FIRST \citep[]{1997ApJ...475..479W}, NVSS \citep[]{1998AJ....115.1693C}, HEASARC, and Sparcl \citep[]{juneau_sparcl_2024}. 

As discussed in \S\ref{sec:pipeline}, some photometry stored in OTTER may be different reductions of the same dataset. This especially affects X-ray and UV data and is intentional because it provides users the freedom to choose their preferred reduction. So, as part of the query implemented in the Python API, we provide an algorithm for finding different reductions of the same data and choosing only one of those photometry points. The algorithm finds any photometry points within a specified date tolerance ($\sim$ 1 day) of the observation (or between the minimum and maximum date for binned data) of the same transient from the same telescope and filter but from a different reference. Then, to choose the photometry to be returned to the user, we first check if only one of the reductions is host subtracted and, if so, we default to that value. If neither reduction or more than one reduction are host subtracted, then we take the most recent reduction. Our testing reveals that this algorithm works very well for the current OTTER dataset. If the user disagrees with this algorithm, they are instead able to ``plug in'' their favorite algorithm to the query (for example, an algorithm that always prefers the reduction from one paper over another).

Since OTTER only stores basic host information, we built in additional methods which allow users to easily access information on the potential host galaxies of these transients. The host galaxies of TDEs are generally unambiguous since TDEs are typically found in the nucleus. However, for other types of transients, we implement a probability of chance coincidence calculator, as defined in \citep{bloom_observed_2002}. Additionally, we built in a method to try to find the host information using the Blast API \citep{jones_blast_2024, gagliano_ghost_2021}, if none exists by default.

In addition to the OTTER Python API, ArangoDB provides a builtin RESTful API as another programmatic access point of the data. This means those who do not use Python can query the OTTER backend database either from the command line or in their programming language of choice. 

\section{Current State of the OTTER Dataset}\label{sec:current-dataset}

\begin{figure*}
    \centering
    \includegraphics[width=0.8\linewidth]{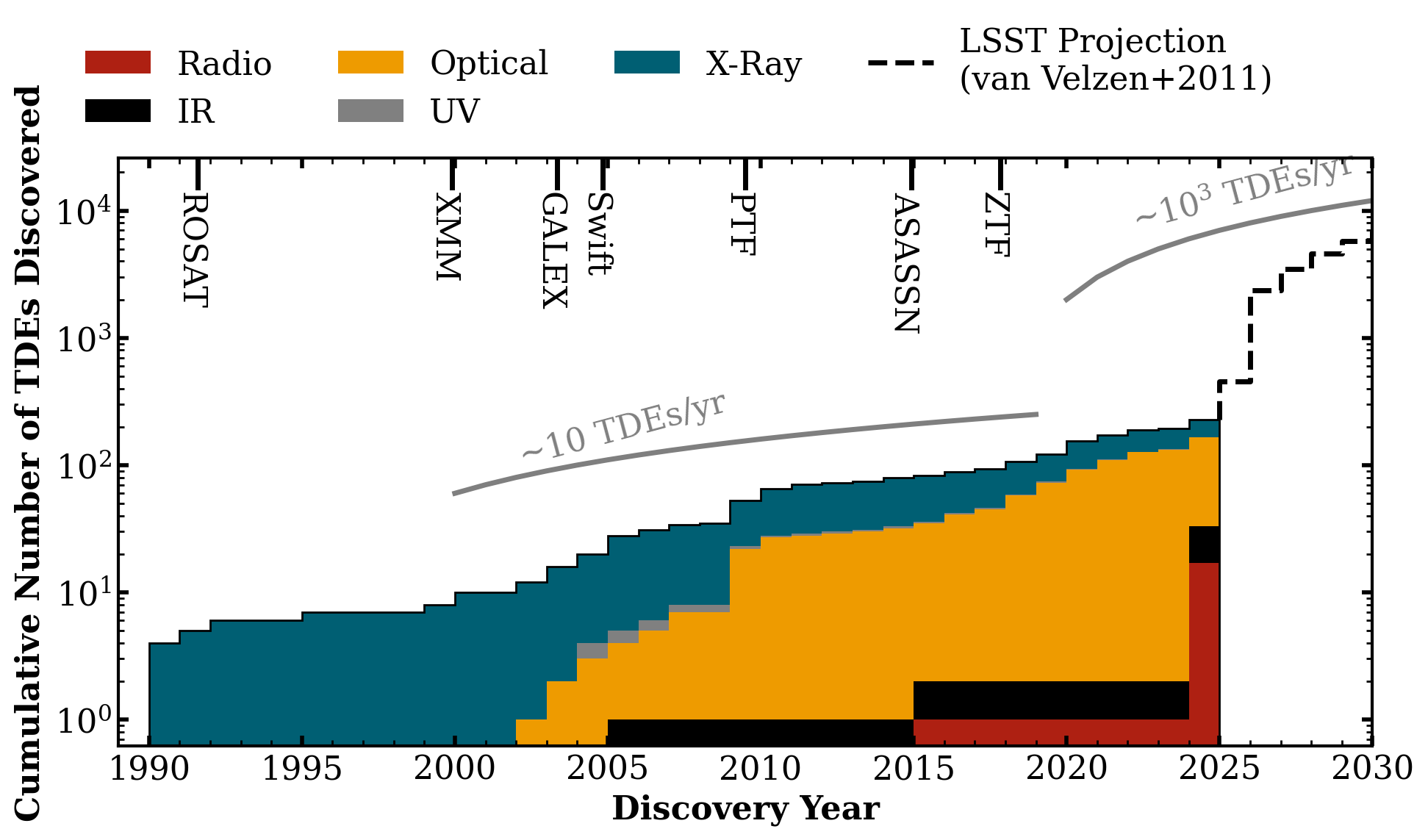}
    \caption{Cumulative histogram of the number of TDE candidates discovered per year, colorized by the wavelength range in which it was discovered. The approximate start date of the major TDE candidate discovery telescopes are labelled on the top. Note the logarithmic scale on the y-axis. \autoref{fig:non-cumulative-disc-hist} is a non-cumulative version of this histogram in \autoref{app:non-cumulative-hist}. A movie of this figure with a linear y-axis is available at: \url{https://github.com/astro-otter/examples/blob/main/tde-discovery-histogram-linear.gif}}  
    \label{fig:tde-discovery}
\end{figure*}

\begin{figure}
    \centering
    \includegraphics[width=\linewidth]{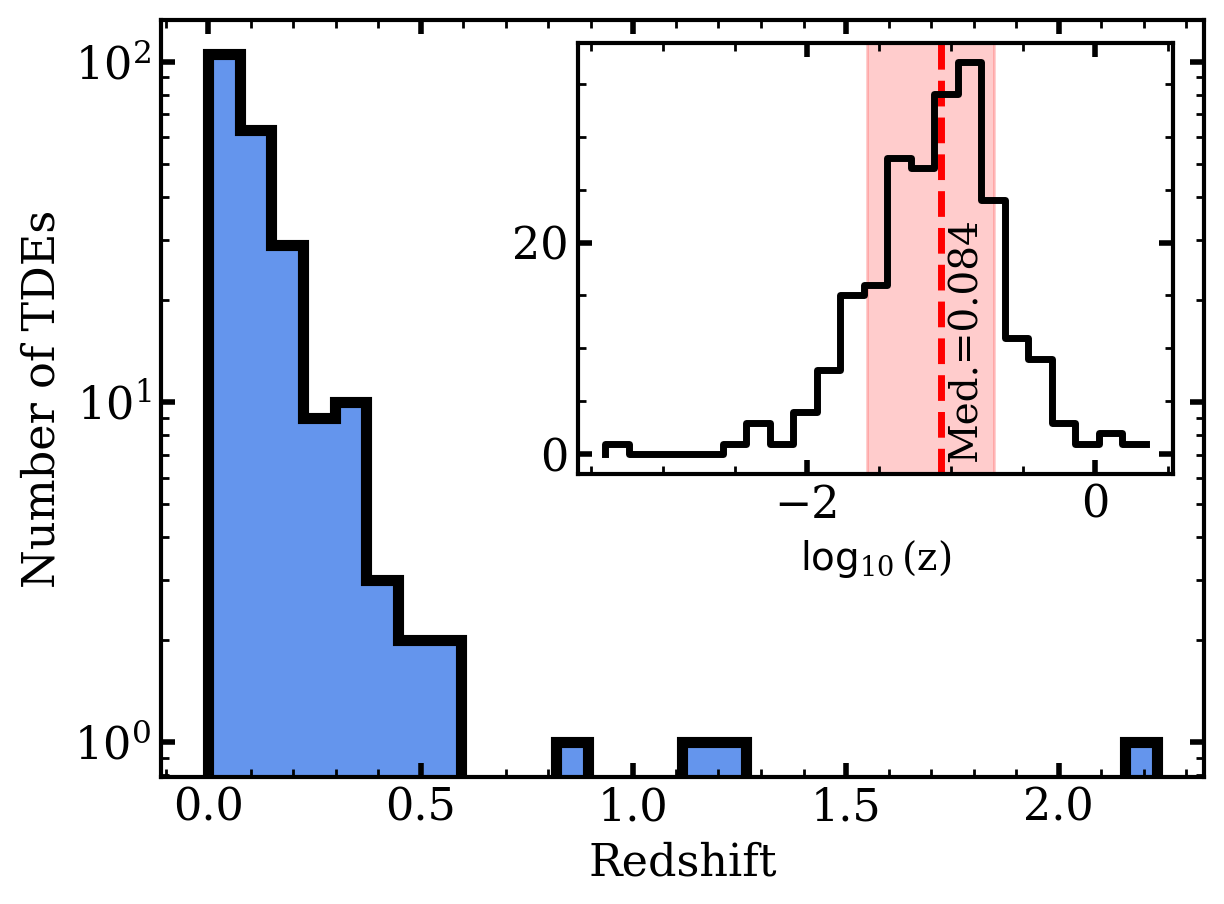}
    \caption{Logarithmic redshift distribution of transients in OTTER. The median of the distribution is around 0.1 and is likely because this is the approximate redshift limit for observing a typical non-jetted TDE with the Zwicky Transient Facility. }
    \label{fig:tde-z}
\end{figure}

\begin{figure}
    \centering
    \includegraphics[width=\linewidth]{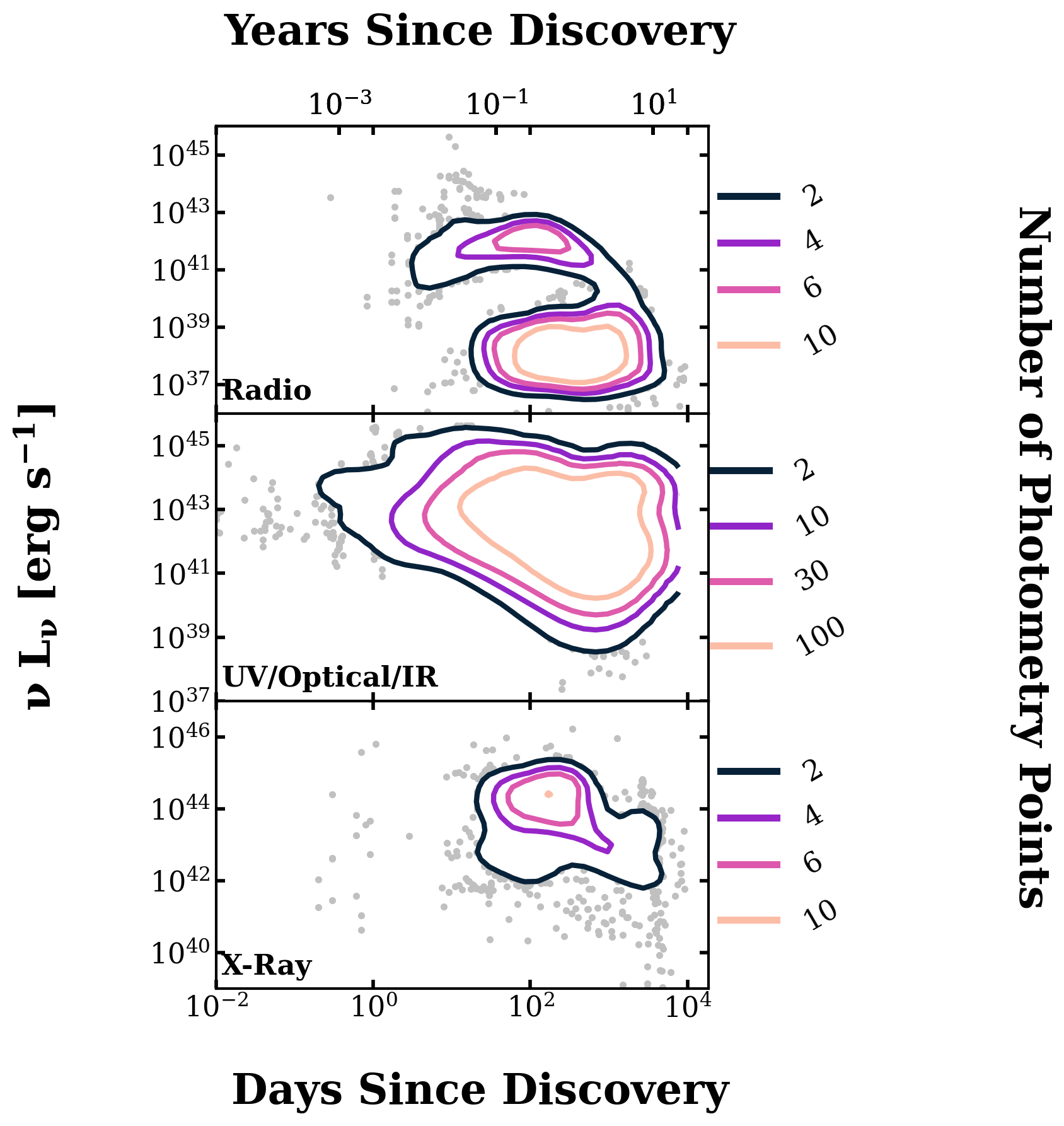}
    \caption{Light curves in $\nu L_\nu$ space of all of the photometry in OTTER. The colorbars have different minima and maxima based on the number of observations in that wavelength regime and are normalized by the width of the bin in days. Note the two distinct classes of Radio TDE (or TDE candidate) light curves (jetted and non-jetted radio emission). The UV/Optical/IR and X-ray photometry span $\sim 6$ orders of magnitude in luminosity space, demonstrating a diversity of in UVOIR and X-ray TDE candidate luminosities.}
    \label{fig:tde-lc}
\end{figure}

\begin{figure*}
    \centering
    \includegraphics[width=\linewidth]{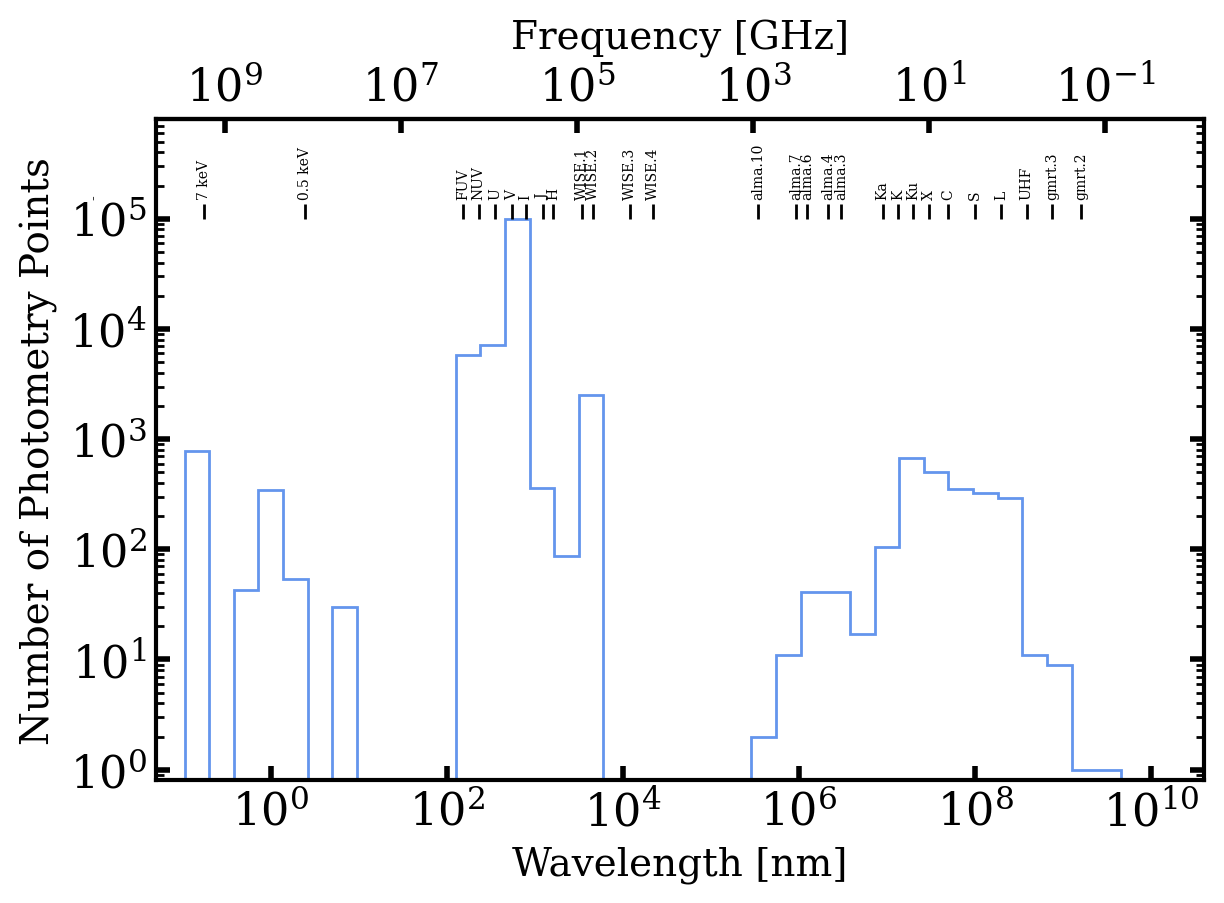}
    \caption{Distribution of photometry points in OTTER as a function of Wavelength/Frequency. We have labelled some typical UVOIR and radio filters/bands along the top for reference. The X-ray energy labels are present as a reference and the individual X-ray photometry points are considered solely based on the center wavelength of the energy range used for each observation. Most of the photometry is in the UVOIR bands and the X-ray and Radio distributions have about the same number of photometry points.}
    \label{fig:tde-phot}
\end{figure*}

As of early December 2025, OTTER contains data on \ntdes TDE candidates, spanning 34 years of observations, making it the largest catalog of TDE candidate photometry and metadata to date. \autoref{fig:tde-discovery} summarizes the TDE candidate discovery rate by different X-ray, radio, and optical surveys over this time period. It also shows the projection for the TDE discovery rate over the next 5 years as the Rubin Observatory transient discovery rate increases. This is predicted to increase the TDE discovery rate by two orders of magnitude \citep[]{van_velzen_optical_2011,bricman_prospects_2020}. 

For every transient in OTTER, we have at least one set of coordinates. $227/\ntdes\approx95\%$ of the transients in OTTER have a spectroscopic redshift. The transients missing a redshift are MAXI J1807+132, which is an X-ray selected TDE candidate, and the 12 radio selected TDE candidates from \citet{dykaar_untargeted_2024}. \autoref{fig:tde-z} shows the redshift distribution of known TDE candidates in OTTER, not corrected for observational bias. The peak appears around $z\sim0.09$ and 68\% of TDE candidates lie in the range from $z\sim0.03 - 0.2$, which is likely the result of an observational bias. 

$203/\ntdes \approx 85\%$ of the transients in OTTER have a discovery date stored. The missing discovery dates are from TDE candidates that were not discovered on a specific date but were rather found in archival searches \citep[e.g.,][]{masterson_new_2024, somalwar_vlass_2023, dykaar_untargeted_2024}. We hesitate to set the discovery date of these events to the first archival detection since, generally, archival searches have much larger gaps between observations than dedicated surveys. This means it is difficult to interpret the date of the first detection as the discovery date under the typical definition. 

\begin{figure}
    \centering
    \includegraphics[width=0.9\linewidth]{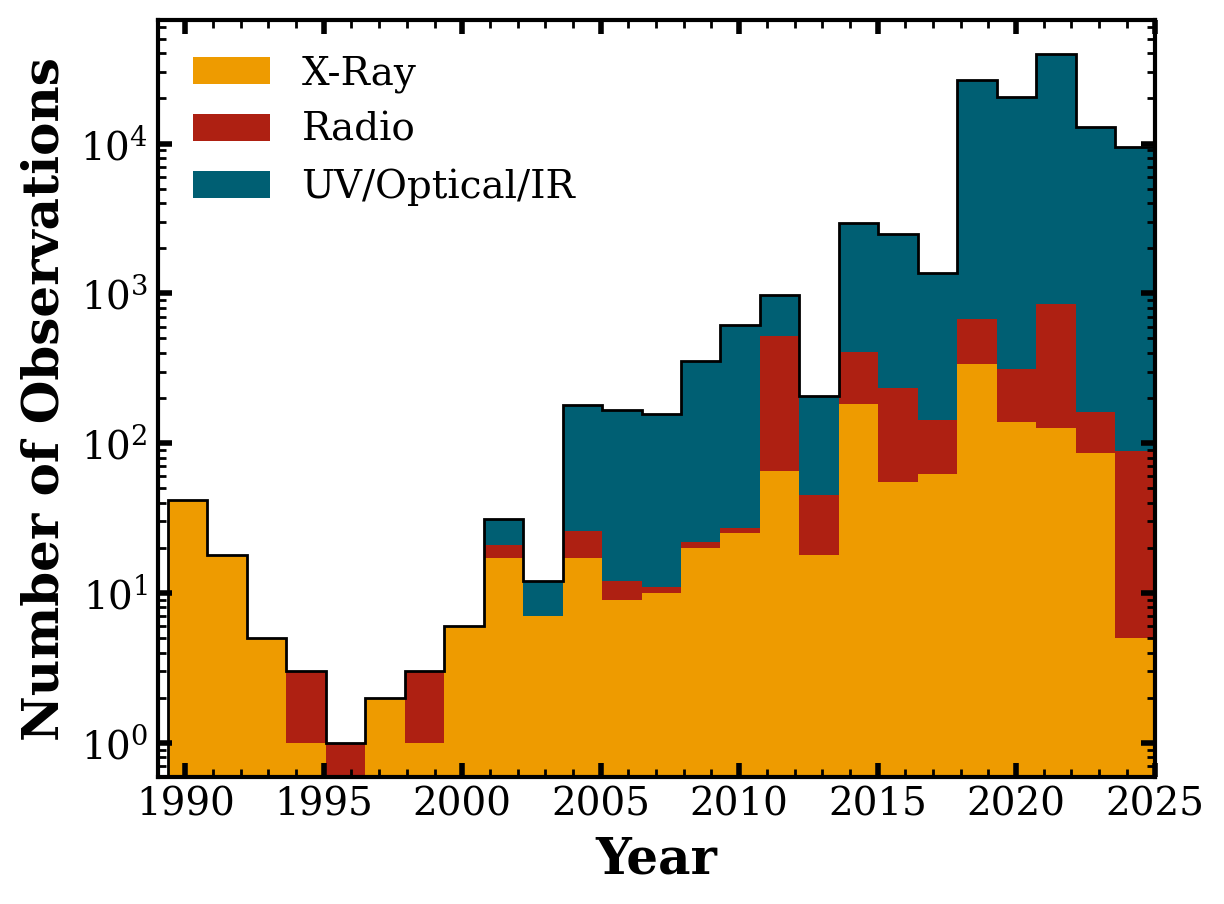}
    \caption{The distribution of photometric observations in OTTER as a function of time. The colors correspond to different observation types: Radio (red), UVOIR (blue), and X-ray (yellow). }
    \label{fig:tde-phot-hist}
\end{figure}

All the photometry in OTTER is shown in Figures \ref{fig:tde-lc}, \ref{fig:tde-phot}, and \ref{fig:tde-phot-hist} and contains $\gtrsim \added{118,000}$ photometry points. $\added{193/\ntdes = 80\%}$ of the TDE candidates in OTTER have at least one photometry point at any wavelength. $\added{98/\ntdes \approx 41\%}$ of the TDE candidates in OTTER have at least one radio observation, $141/\ntdes \approx 59\%$ of TDE candidates in OTTER have at least one UVOIR observation, and $82/\ntdes \approx 34\%$ of TDE candidates in OTTER have at least one X-ray observation. Some of the missing photometry simply demonstrates inconsistencies in the multiwavelength follow up of TDE candidates. In more recent years, more TDE candidates were followed up more consistently across the electromagnetic spectrum (e.g., \autoref{fig:tde-phot-hist}). 

Note that in years since $\sim 2022$, our photometric completeness begins to fall off and we have less photometry after $2024$. This is shown in Figure \ref{fig:tde-phot-hist} and is expected because OTTER only contains {\it published} photometry, so the more recently discovered TDE candidates simply have fewer published observations.

\section{Example Applications of OTTER}\label{sec:sample}

The OTTER infrastructure has a wide variety of potential applications across the field of transient astronomy, ranging from a quick view of existing public datasets to detailed quantitative comparison with models and locally-stored datasets. Here, we list a few brief examples of applications of the catalog and infrastructure to the TDE candidate population currently stored in OTTER. We provide jupyter notebooks as part of the documentation demonstrating each of these analyses.

\subsection{Example 1: Quantifying the Delayed X-ray Emission in Optically-Discovered TDEs}
Here, we focus on testing the two competing models introduced in \S\ref{sec:intro} for the UV/Optical emission observed from TDEs: the stream-stream and reprocessing models. While the processes described in both models could be occurring, the UV/optical emission is likely dominated by only one \citep[]{gezari_tidal_2021}. Both of these models predict that the X-ray emission lags behind the UV/Optical emission. However, the predicted time delay differs between models, making quantifying the time delay a potentially interesting observational test. In the stream-stream model, the time delay between the emission components is caused by the ``circularization time'', i.e. the time it takes the material to circularize into an X-ray-producing accretion disk after producing UV/Optical emission during the stream-stream collision. Theoretical predictions of the circularization time vary by about an order of magnitude, from $\sim 50-70$ days \citep[][their Fig. 2]{steinberg_streamdisk_2024}\footnote{Note, however, that \citet{steinberg_streamdisk_2024} finds that the time of peak optical and X-ray light is roughly comparable.} to $\sim$years (\citealt{shiokawa_general_2015}, their eqn. 11 \& 12;  \citealt{piran_disk_2015}, their eqn. 4 and the subsequent text with corrections for circularization). In the reprocessing model, the time delay is the time it takes for the intervening material to dissipate and reveal the X-ray-producing accretion disk. The dissipation time is predicted to be $\sim 100-200$ days but is dependent on the properties of the system \citep[e.g., SMBH mass, stellar mass, envelope opacity, accretion rate, etc.;][eqn. 33 \& 34]{metzger_cooling_2022}.  

\newcommand{\fw}{0.8}
\begin{figure*}[!t]
\begin{center}
\subfloat[]{
  \includegraphics[width=\fw\linewidth]{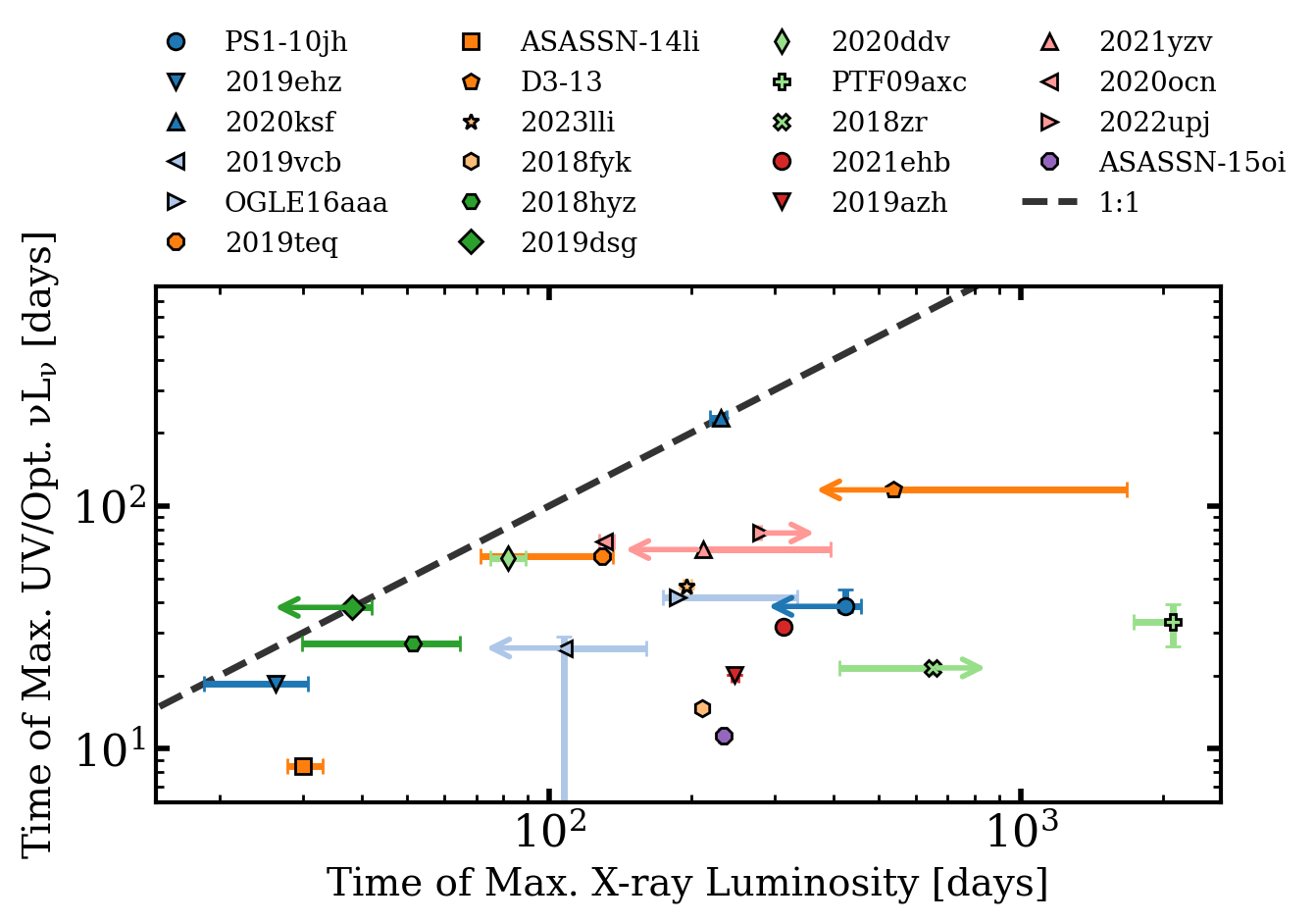}
  \label{fig:max-xray-uvoir}
}
\hfill
\subfloat[]{
  \includegraphics[width=\fw\linewidth]{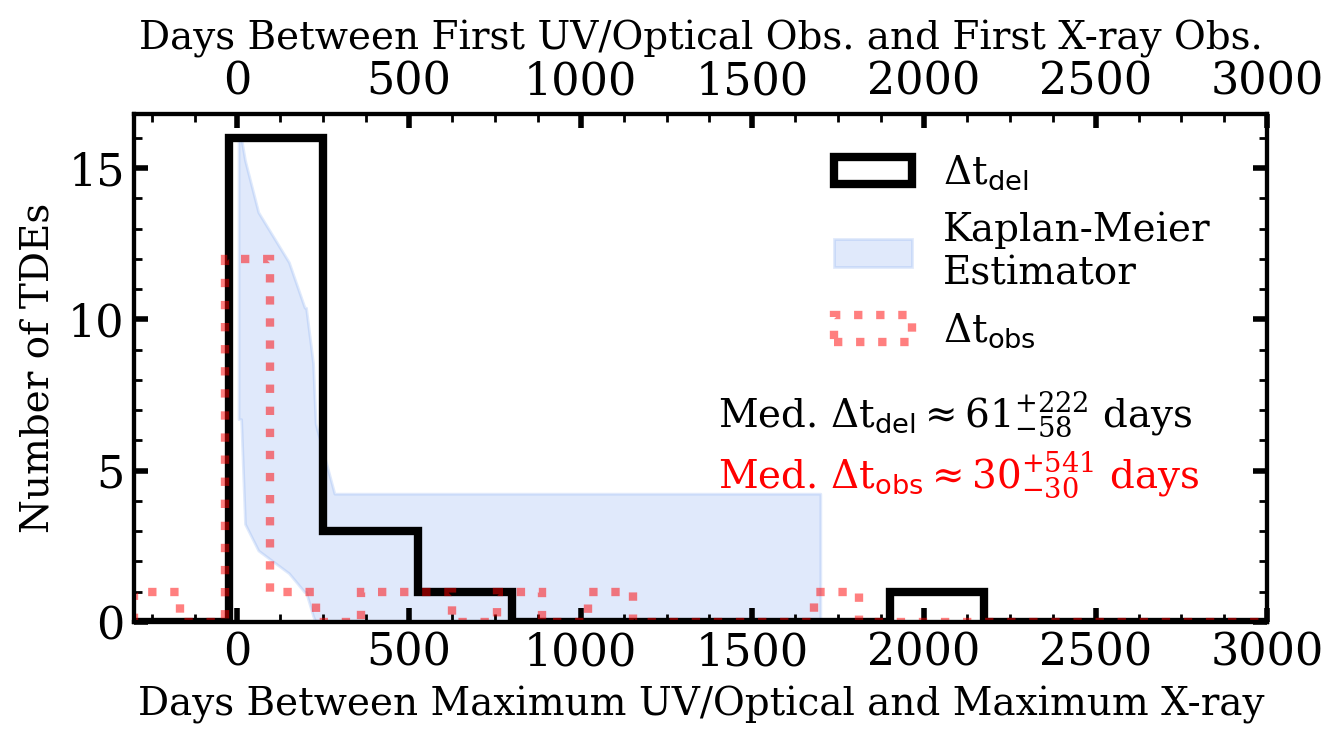}
  \label{fig:max-xray-uvoir-hist}
}
\end{center}
\caption{{\it Figure \ref{fig:max-xray-uvoir}:} The days since TDE discovery date for the maximum UV/Optical luminosity vs X-ray luminosity for a sample of optically-selected TDEs. Most points lie either on or below the grey dashed 1:1 line, indicating that most of the population has a delay between the maximum of the UV/Optical emission and the maximum of the X-ray emission. {\it Figure \ref{fig:max-xray-uvoir-hist}:} A histogram of the days between the maximum UV/Optical and X-ray luminosity ($\Delta t_{\rm del} \equiv \Delta t ({\rm X-ray}) - \Delta t({\rm UVOIR})$, where $\Delta t$ is the days since discovery). The blue line shows the Kaplan-Meier fit to the distribution, which considers the upper and lower limits, and the blue region shows the 1$\sigma$ error on that fit. The outlier at a large delay time of $\sim 2000$ days is PTF09axc, which was optically discovered in 2010 and had no X-ray detection (but does have X-ray upperlimits) until 2014. For comparison, we also show a histogram of $\Delta t_{\rm obs}$, the time between the first optical and the first X-ray observations of TDEs in OTTER.}
\label{fig:xray-uvoir-timing}
\end{figure*}

\citet{guolo_systematic_2024} found, qualitatively, that the X-ray emission of a subset of optically-selected TDEs is delayed compared to their UV/Optical emission. To further explore this, we quantify the time delay for a large sample of TDEs. We select all of the TDEs in OTTER (as of November 2025) that have \added{{\tt unambiguous=True}, at least one detection, and at least two observations in both the X-rays {\it and} UV/Optical wavelengths}, with a few exceptions. We exclude the jetted TDEs Sw J1644+57, Sw J2058+05, and AT\,2022cmc since the processes producing the UV/Optical and X-ray emission are unlikely to be the same as the rest of the TDEs in this sample \citep[]{2011Natur.476..425Z, 2011Sci...333..199L, 2012ApJ...748...36B, 2012MNRAS.421.1942W, 2013ApJ...767..152Z, 2016ApJ...819...51L, 2016MNRAS.462L..66Y, 2017ApJ...838..149A, 2018ApJ...854...86E, 2021ApJ...908..125C,2012ApJ...753...77C,2015ApJ...805...68P,2017ApJ...838..149A,2017MNRAS.472.4469B}. \added{We also exclude any events discovered in archival searches since the discovery date is arbitrary and we therefore are unable to compute $\Delta t ({\rm X-ray})$ and $\Delta t({\rm UVOIR}$)}. Note, we do not consider infrared photometry for this analysis ($\lambda_{\rm eff} \geq 1 ~{\rm \mu m}$) because the infrared emission, which is often dominated by dust echoes, appears to be unrelated to the early-time UV/Optical emission \citep[]{van_velzen_reverberation_2021, van_velzen_first_2023}. These cuts resulted in a sample that is entirely optically selected. 

\newcommand{\delaytime}{$\Delta t_{\rm del}$\,}
\newcommand{\uvotime}{$t_{\rm UVO}$\,}
\newcommand{\xtime}{$t_{\rm X}$\,}
After selecting the sample, we find the days since discovery at the maximum point of both the observed UV/Optical (\uvotime) and X-ray (\xtime) lightcurves for each TDE. However, the time corresponding to these flux maxima may not be the true lightcurve maximum. To account for this we use the time between the two points on either side of the observed maxima as uncertainties on the observed lightcurve maxima. A scatter plot of these dates is shown in \autoref{fig:max-xray-uvoir}. To quantify the time delay (\delaytime), we then subtract \uvotime from \xtime. We propagate the uncertainties on the date of maximum flux by finding both the minimum and maximum possible delay times. Then, for later statistical analysis, we use this minimum and maximum possible delay time as the bounds of an interval. The distribution of \delaytime is shown in \autoref{fig:max-xray-uvoir-hist}. We note that \autoref{fig:max-xray-uvoir-hist} shows a skewed unimodal distribution with a tail towards large positive \delaytime. 

Since our sample only includes optically discovered TDEs that were later followed up with X-ray observations, there may be a systematic bias of \delaytime to positive values greater than the true population \delaytime distribution. We partially account for this bias by introducing upper limits when the maximum was the first observation and lower limits when the maximum was the last observation (e.g., see the arrows in \autoref{fig:max-xray-uvoir}). For subsequent statistical analysis, if \xtime is an upper limit or \uvotime is a lower limit then \delaytime is treated as a lower limit. In contrast, if \xtime is a lower limit or \uvotime is an upper limit then \delaytime is treated as an upper limit. 

To further quantify the impact of the bias introduced by an optically-selected sample, we calculate the distribution of the lag time between the first UV/Optical observation and the first X-ray observation ($\Delta t_{\rm obs}$; red dashed line in \autoref{fig:max-xray-uvoir-hist}). 
We then compare the $\Delta t_{\rm obs}$ and \delaytime distributions using a Kolmogorov-Smirnov test and find that we cannot rule out that the \delaytime distribution is affected by the aforementioned sample selection bias. Therefore, since the \delaytime is likely shifted to higher positive values by this observational bias, the \delaytime distribution, and any related summary statistics, should be treated as upper limits. 

To account for the uncertainties and limits on the delay times in our statistical analysis, we fit a Kaplan-Meier estimator \citep[]{Kaplan01061958} to the \delaytime distribution\footnote{To be more specific, we fit the Kaplan-Meier estimator using the uncertainties and/or limits as interval censored data, essentially treating each of these \delaytime values as a range (which could have a lower limit of $-\infty$ or an upper limit of $+\infty$).}. The best-fit Kaplan-Meier estimator has a median \delaytime of $61$ with 16th and 84th quartiles of 3 and 282 days, respectively. We additionally compute a $95\%$ Greenwood confidence interval \citep[]{19272700028} and find that the true population median of \delaytime lies in the range $3-228$~days. 

Another potential bias, which may also change \delaytime, occurs because the available X-ray observations for different TDEs are not be equally sensitive. For example, some TDEs have only Swift follow up, but a typical Swift 2ksec observation is insufficient to detect some of the fainter X-ray emission detected with, e.g., XMM-Newton or Chandra \citep[e.g., ][]{guolo_systematic_2024}. This may result in poor estimates for some of the \xtime measurements due to a lack of deep X-ray limits. We attempt to account for this by using uncertainties and limits on the measurement of the maximum time, but nevertheless this bias may reduce our constraining power from the data available. Future observations with Einstein Probe \citep[]{yuan_einstein_2022} will likely be able to reduce this bias and improve the constraining power.

Overall, based on the quartiles of the \delaytime distribution, we find a \textit{conservative} upper limit $\Delta t_{\rm del} \lesssim 300$ days, with one exceptional event with an extremely long delay time (PTF09axc\footnote{\url{https://otter.idies.jhu.edu/transient/PTF09axc}}). This indicates that, for most TDEs, we disfavor, but cannot completely rule out, theoretical stream-stream models that predict longer circularization times than a few hundred days. However, as discussed above, there are numerous biases that are difficult to fully account for. Therefore, the data may still be consistent with many of the previously presented models \citep[]{piran_disk_2015, shiokawa_general_2015, dai_unified_2018, metzger_cooling_2022, steinberg_streamdisk_2024}. 

If the \delaytime distribution is interpreted as the circularization time, the wide spread we find in \delaytime could be further evidence that the circularization time is physically related to the initial period of the stellar orbit \citep[]{bonnerot_formation_2021}, SMBH properties \citep[e.g., mass and spin, ][]{steinberg_streamdisk_2024}, or nodal precession of the plane of the stellar stream \citep[]{guillochon_hydrodynamical_2013, bonnerot_formation_2020}. If, instead, the time delay is interpreted as the dissipation time of a reprocessing envelope, the spread in the distribution can be interpreted as a viewing angle effect \citep[]{dai_unified_2018, metzger_cooling_2022}. Moving forward, 1) a more explicit theoretical prediction of this \delaytime distribution and 2) deeper early-time X-ray observations of TDEs may be necessary to distinguish between them. We defer further investigation of these potential relationships to future work.

\subsection{Example 2: The Dusty Environments of TDE Extreme Coronal Line Emitters}
Extreme Coronal Line (ECL) Emitters (ECLEs)\footnote{Such events are typically selected based on strong Iron lines (e.g., [Fe X]), which were first observed in the solar corona, hence the name ``coronal lines''.} are galaxies observed to emit strong, high-energy ionization metal lines \citep[e.g.,][]{komossa_discovery_2008, wang_extreme_2012}. Such high-ionization metal lines require $\sim$300 eV radiation and, therefore, require at least a soft X-ray source \citep[]{mazzalay_demystifying_2010}. One such source of this X-ray radiation is an accretion disk surrounding a SMBH, like those present in AGN and TDEs, indicating that theoretically a large fraction of TDE hosts should exhibit ECLs post disruption. However, curiously, this is not the case: Only a small fraction of TDE hosts exhibit ECLs \citep[]{yang_long-term_2013, clark_long-term_2023, callow_rate_2024, onori_nuclear_2022, somalwar_first_2023, hinkle_coronal_2024, newsome_mapping_2024, koljonen_extreme_2024, clark_at_2025}. Recent work suggested that this discrepancy is due to differences in the galaxy nuclei gas density of ``typical'' TDE host galaxies and the host galaxies of TDEs that produce ECLs \citep[]{hinkle_coronal_2024, mummery_galaxy_2025}. We explore this relationship further by combining radio data with infrared data of a population of ECLEs, both of which provide independent probes of the  properties of the circumnuclear medium.

OTTER has the ability to assist with such analyses in multiple ways. First, many of the optically selected TDEs that exhibit ECLs, like TDE\,2019qiz \citep[]{nicholl_outflow_2020, short_delayed_2023}, already have detailed previously-published multiwavelength datasets stored in OTTER. Second, the OTTER API has numerous methods for easily scraping other existing catalogs for even more detailed datasets of the transient events. These include publicly available datasets provided by WISE, ZTF, ASASSN, etc. Third, the OTTER Python API can easily handle combining local data with public OTTER datasets, allowing for simultaneous queries of the two data sources with the same useful de-duplication code running that is used for cleaning a new dataset on upload.  

\begin{figure}
    \centering
    \includegraphics[width=0.9\linewidth]{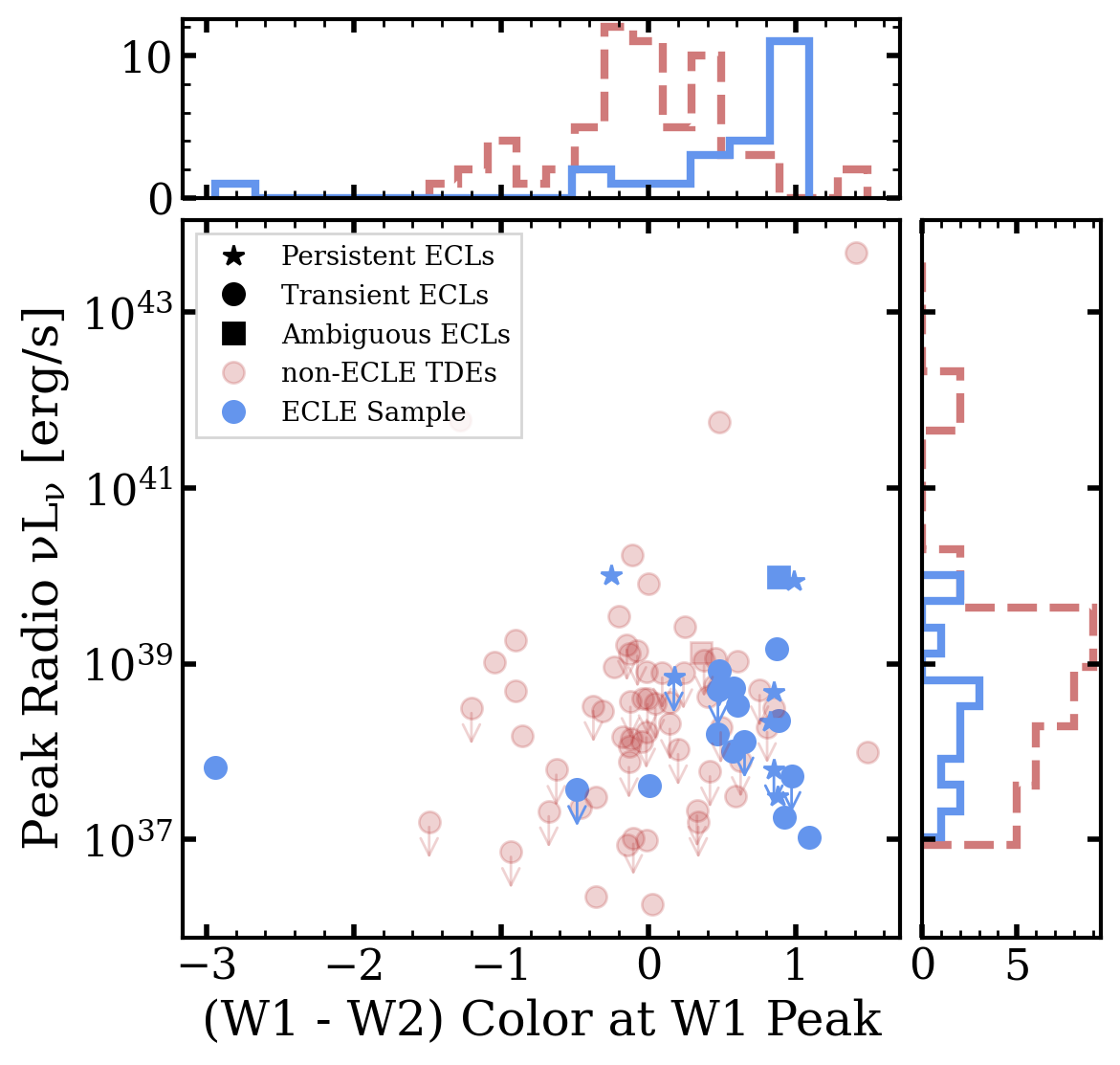}
    \caption{The peak radio luminosity vs the wise color (W1-W2) at the maximum W1 value. The blue points are the ECLE sample while the red points are the broader TDE population, which do not show evidence of ECLs. In the histograms, we show the ECLE sample as a solid blue line 
    and the broader TDE population as a dashed red line. The marker style shows the literature classification of each of the ECLEs. The TDE-ECLE dataset includes private radio data (Franz, N et al., in prep) while the TDE sample includes radio data already present in OTTER. All of the WISE data is produced from querying the WISE public archive \citep[]{2020MNRAS.493.2271H, WISE} using the OTTER API and is not host subtracted.}
    \label{fig:ecle-radio-wise}
\end{figure}

As a sample analysis, we plot the peak radio luminosity (a mix of public and private data; Franz, N et al., in prep) against the WISE colors (which we obtain by using the OTTER API to query WISE) for the ECLE-TDEs in \autoref{fig:ecle-radio-wise} and compare these to the non-ECLE-TDEs population. Using the OTTER dataset from November 2025, we find that ECLE-TDEs tend to occur in environments with redder dust than the broader TDE population with a high statistical significance ($p \sim 10^{-7}$, one-sided Kolmogorov-Smirnov test; $p \sim 10^{-5}$, one-sided Wilcoxon rank-sum test). By extension, this may indicate that ECLE-TDEs occur in environments with higher density dust\footnote{There is also the possibility that this difference instead indicates a different amount of incident flux onto the dust grains, especially in the case of the archival ECLEs that do not have constraints on the start of the observed emission. But, consider that \citet{hinkle_coronal_2024} found that the WISE {\it color} is also transient in a subset of our ECLE sample, indicating that the dust heating is transient in these events. We defer further discussion disentangling these scenarios to future work as it will likely require a more detailed multiwavelength investigation.}. This is a similar finding to \citet{hinkle_coronal_2024} but with a larger sample of events. The radio data in this figure will be discussed in more detail in Franz, N et al., in prep.

\subsection{Example 3: Easily Modeling TDE Light-Curves}
Another advantage of consolidating the TDE candidate data is that it makes it very easy to apply existing modeling codes, such as MOSFiT \citep{guillochon_mosfit_2018, mockler_weighing_2019}, \texttt{redback} \citep[]{sarin_redback_2024}, \texttt{FitTed} \citep[]{mummery_fitting_2024}, and \texttt{TDEMass} \citep[]{ryu_measuring_2020, ryu_tdemass_2020}. By fitting these models to the data we can derive properties of the SMBH, progenitor star, and cirumnuclear environment. Most notably, these models produce a galaxy-independent measure of the SMBH mass which is either constrained by the light curve as a whole (e.g., MOSFiT) or from the late-time plateau of the light curve (e.g., FitTed). 


\newcommand{\fww}{0.48}
\begin{figure*}[!t]
\begin{center}
\subfloat[]{
  \includegraphics[width=\fww\textwidth]{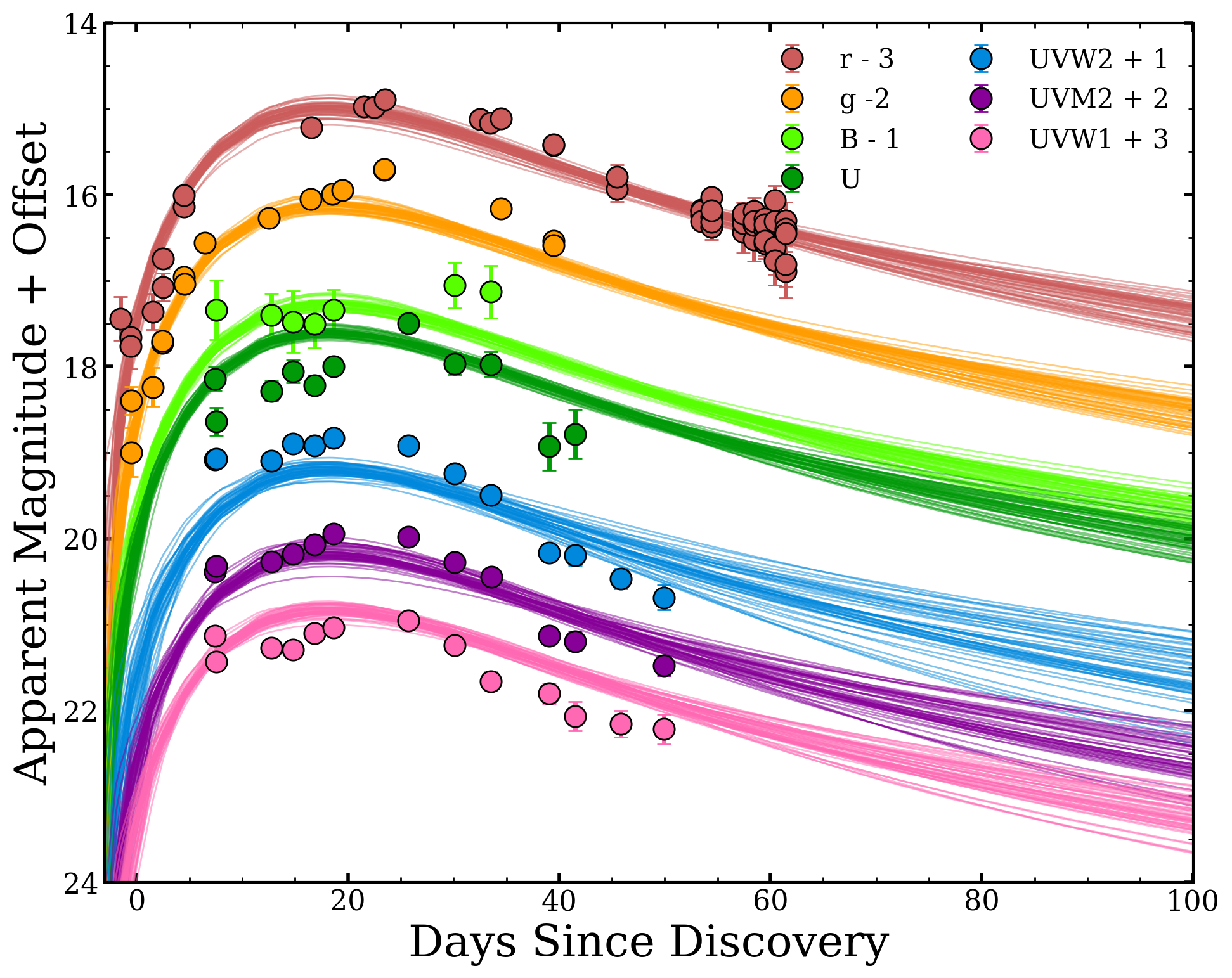}
  \label{fig:20zso_mosfit_model}
}
\subfloat[]{
  \includegraphics[width=\fww\textwidth]{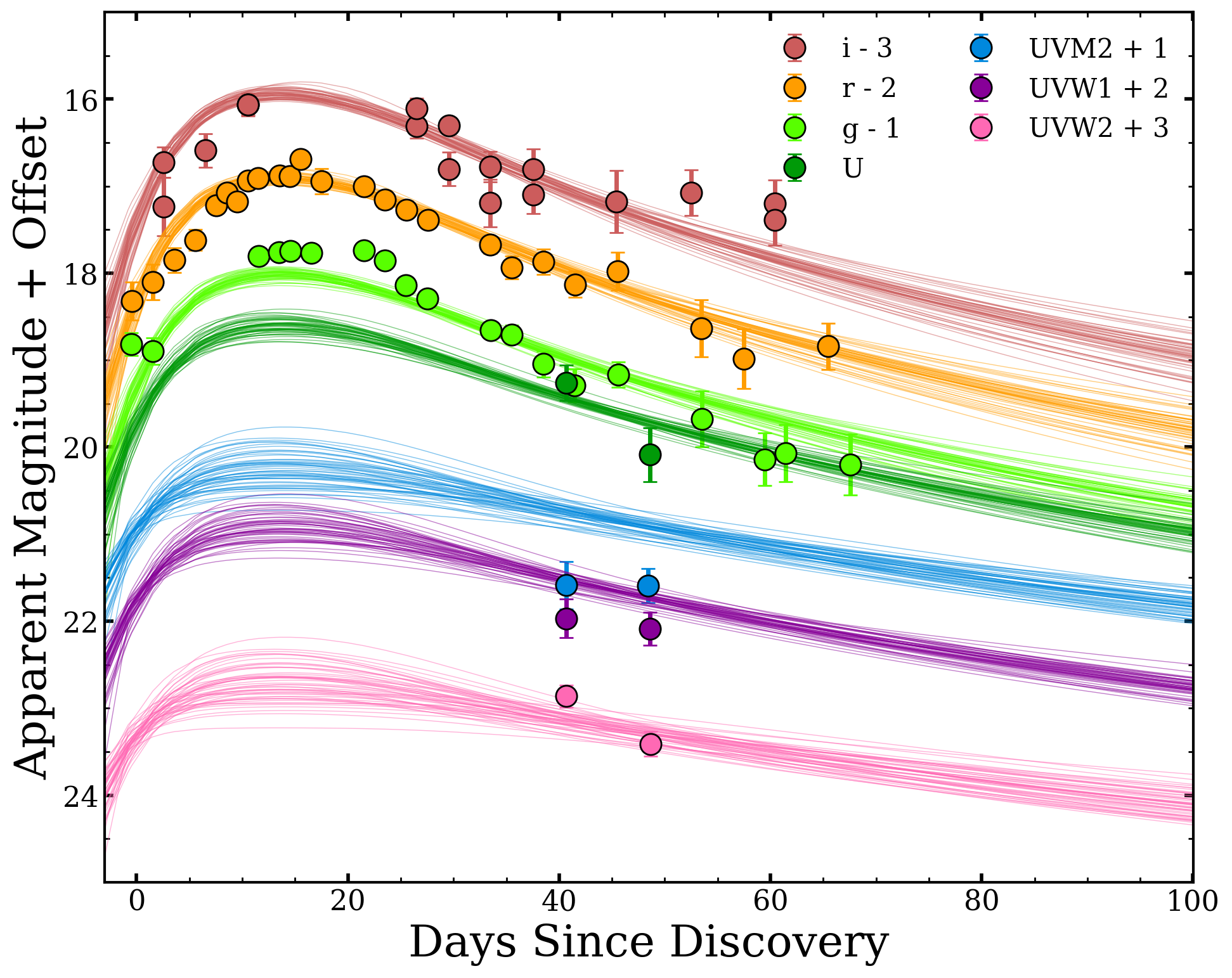}
  \label{fig:21sdu_mosfit_model}
}
\end{center}
\caption{Example MOSFiT models of TDE\,2020zso (left) and TDE\,2021sdu (right). By eye, the MOSFiT models generally appear to fit the light curves well. From these fits we find that the mass of the SMBH in the host galaxy of TDE\,2020zso is $\log_{10}\left(M_{BH}\right) =  6.14^{+0.13}_{-0.24}$ and of the TDE\,2021sdu host galaxy is $\log_{10}\left(M_{BH}\right) = 5.53^{+0.22}_{-0.25}$.}
\label{fig:mosfit-ex}
\end{figure*}

In this work, we focus on applying MOSFiT to some of the objects in the OTTER dataset. The MOSFiT TDE model begins after the circularization ceases, when accretion onto the SMBH is actively occuring. MOSFiT then assumes that the accretion disk produces thermal X-rays which are reprocessed into the UV/Optical. As an example, we can model two TDEs recently discovered by ZTF: TDE\,2020zso \citep{2020TNSTR3449....1F, 2020TNSCR3486....1I, hammerstein_final_2023, 2024MNRAS.527.2452M} and TDE\,2021sdu \citep{2021TNSTR2318....1M, 2021TNSCR2712....1C, yao_tidal_2023, 2024MNRAS.527.2452M}. We show these models in \autoref{fig:mosfit-ex}. From MOSFiT we find that the mass of the SMBH (mass of the disrupted star) in TDE\,2020zso is $\log_{10}\left(M_{BH}\right) =  6.14^{+0.13}_{-0.24}$ ($\log_{10}\left(M_*\right) = 0.002^{+0.19}_{-0.08}M_\odot$) and of TDE\,2021sdu is $\log_{10}\left(M_{BH}\right) = 5.53^{+0.22}_{-0.25}$ ($\log_{10}\left(M_*\right) = 0.02^{+0.42}_{-0.18} M_\odot$), where the lower and upper uncertainties are the 5\% and 95\% quantiles, respectively. A jupyter notebook demonstrating how to model these two datasets with MOSFiT is available on the OTTER API documentation webpage and shows that it takes $\lesssim100$ lines of code to reformat the OTTER data into an input file for MOSFiT.

\subsection{Example 4: Validating Optical Transient Classifiers in the Rubin era}

As we enter the era of Rubin Observatory's Legacy Survey of Space and Time \citep[][]{ivezic_lsst_2019}, the number of transients discovered will outnumber the amount of telescope time necessary to spectroscopically classify them by a factor of 1000 (e.g., \autoref{fig:tde-discovery}). As a result, alternative methods of transient classification are necessary. One method that has gained popularity in recent years is using machine learning classification on transient photometry \citep[e.g., ][]{villar_supernova_2019,villar_superraenn_2020,villar_deep-learning_2021,gomez_fleet_2020, gomez_identifying_2023, stein_texttttdescore_2023, de_soto_superphot_2024, boesky_splash_2025}. 

A sample application of the OTTER dataset is to apply a machine learning classifier to the available optical data to further test the ability of the classifier to recover the TDE classification from a sample of spectroscopically classified TDEs. We use the Finding Luminous and Exotic Extragalactic Transients (FLEET) algorithm, which is able to classify transients based on their optical light curves as SLSN, TDE, and ``other'' \citep{gomez_fleet_2020, gomez_identifying_2023}\footnote{Note, while we report other classifications here, the classifications besides TDE and SLSN are not validated.}. FLEET requires at least 4 detections in $g$-band or $r$-band, and is able to work with SDSS/ZTF filters (i.e. \textit{ugrizy}) so we select all transients from the OTTER dataset (as of August 2025) that have enough photometry in these filters to model and run FLEET. This results in 83 transients, all of which are spectroscopically classified as TDEs. 

\begin{figure*}
    \centering
    \includegraphics[width=0.9\linewidth]{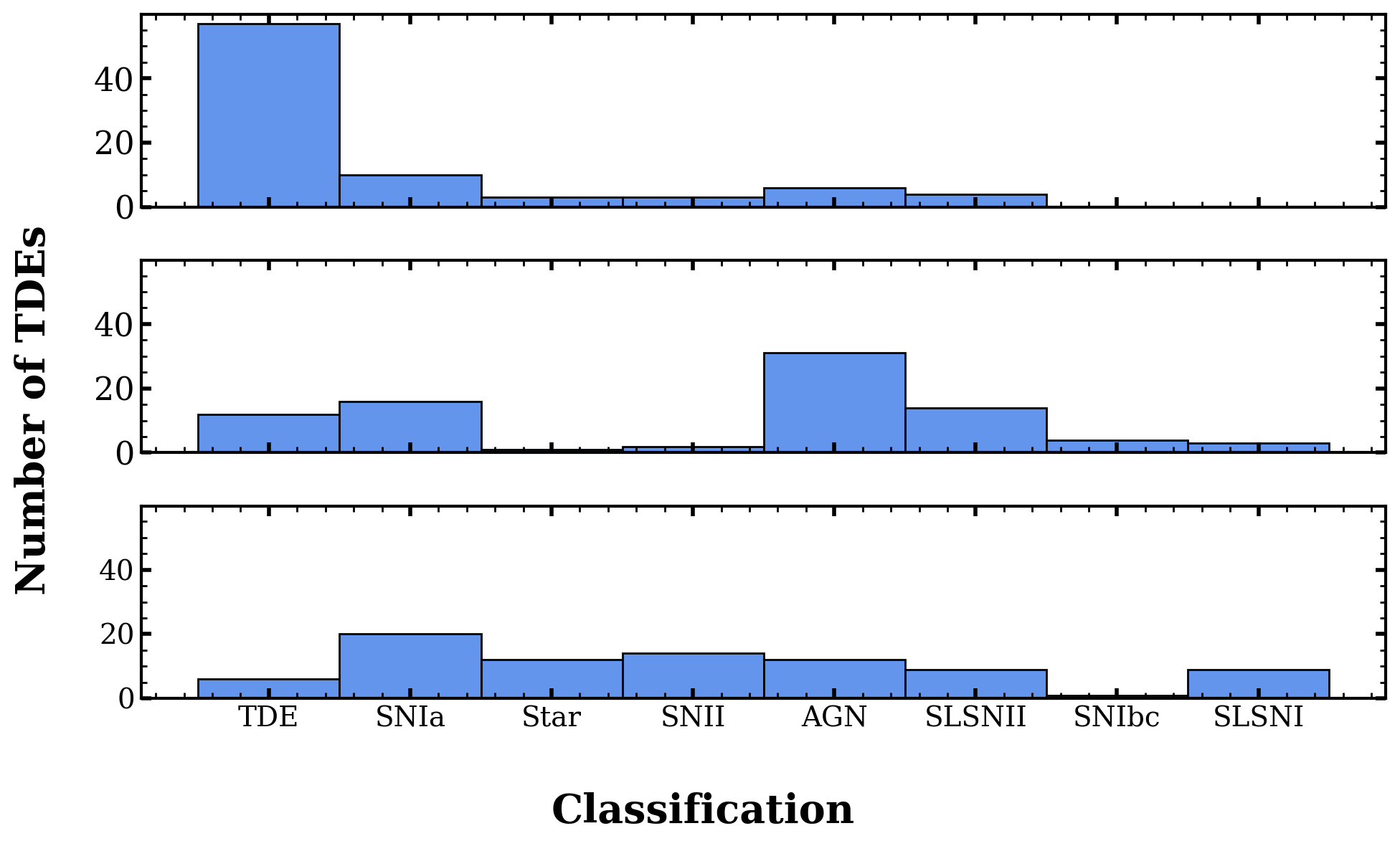}
    \caption{Histograms showing the most probable (top), second most probable (middle), and third most probable (bottom) photometric classification returned by FLEET for a sample of 83 spectroscopically classified TDEs. Generally, the most probable classification returned by FLEET is ``TDE''. In the middle row we see that the other most probable classifications are AGN, Type Ia SNe, and SLSN.}
    \label{fig:fleet-results}
\end{figure*}

The results of this analysis are presented in Figure \ref{fig:fleet-results}, which shows histograms of the top 3 most probable classifications given by FLEET. FLEET finds that the most likely classification is ``TDE'' for $69\%$ of the 83 TDEs. $89\%$ of the 83 TDEs are classified as a TDE as one of the top 3 classifications. Looking at just the most probable classification given by FLEET (top row of \autoref{fig:fleet-results}) we see that the second most common classification found by FLEET for the TDEs is a thermonuclear supernova (SN Ia). We speculate that this is because thermonuclear supernovae are very common\footnote{Even though \citet{gomez_fleet_2020} balances their dataset using resampling techniques, SN Ia did make up 800/1813 transients in the FLEET training sample, which may also bias the results towards SN Ia.} and have somewhat similar rise times to TDEs. 

Looking at the second and third order classifications by FLEET (the bottom two distributions in \autoref{fig:fleet-results}), the next most probable classifications of TDEs are AGN and SN Ia. However, we also see a surge in the number of SLSN II classifications. This is somewhat unsurprising since the luminosities of SLSN II and TDEs are comparable. By the third order distribution, the third most probable classification given by FLEET is essentially a uniform distribution. 

A final notable result is that FLEET appears to fail on the two TDEs in this sample that show on-axis jets (Sw\,J1644+57 and AT\,2022cmc), returning P(TDE)=0\% and finding very low probabilities for all classifications that FLEET is trained on. This is unsurprising given that these TDEs tend to appear much different photometrically from the non-jetted population (although they may share some similarities with the most luminous optical TDEs; \citealt{yao_2025}). Additionally, given the extremely limited sample size of jetted TDEs, FLEET did not include any in their training sample. This may indicate that future TDE classifiers should attempt to classify jetted and non-jetted TDEs separately. In the future, we hope that OTTER will become a hub for curating training and validation datasets for future machine learning classifiers, particularly those that utilize multiwavelength datasets.

\section{Conclusions}\label{sec:conclusion}

We present the Open mulTiwavelength Transient Event Repository (OTTER). OTTER is an open-source, publicly accessible, and scalable infrastructure and catalog for archiving transient metadata and photometry. For the first time, we include photometry from across the electromagnetic spectrum, from radio to X-rays, of transient events. The dataset is stored in the non-relational document database ArangoDB, which allows for maximum data flexibility while still allowing us to outline a detailed schema. This infrastructure also includes two APIs (Python-based and RESTful), with multiple available endpoints, and a web application for interfacing with the dataset. The web-application is available at \url{https://otter.idies.jhu.edu} and the API documentation is available at \url{https://astro-otter.readthedocs.io}, which includes numerous tutorials describing how to access the dataset. 

Both the API and web-interface allow for querying the catalog based on commonly used metadata like (partial) name, coordinates, redshift, classification, and classification flags (i.e., is the classification based on an optical spectrum). The API includes code to easily access both the metadata and the photometry associated with each transient. The web-application has tools for both accessing relevant citations and uploading data, which we encourage other astronomers to include in their workflow once their research has been published. The dataset is currently backed up weekly to GitHub and, if the dataset size becomes too large, we plan to transition to backing it up to Zenodo in the future.

For the initial release of this infrastructure, we also present the largest catalog of tidal disruption event candidates (TDE candidates). As of November 2025, the catalog includes \added{$\sim 115,000$ UVOIR, $\sim 1,250$ X-ray, and $\sim 2,420$ radio photometry points} from \ntdes TDE candidates that were classified in at least one published source. The API and web-application allow for easy access to this dataset, and the OTTER upload form will allow others to upload their datasets after publishing so that, as a field, we can continue to maintain this catalog. 

We present four sample analyses using the available TDE data including a delay time analysis, a sub-population analysis, simple modeling, and validating a machine learning classifier. First, we perform a simple analysis of the TDE population and compute the delay time between the peak UV/Optical and X-ray emission to likely be $\lesssim 300$ days for the population. Second, we briefly discuss an analysis of the population of known Extreme Coronal Line Emitters (ECLEs) in the context of the broader TDE population. We recover the apparent relation that ECLEs appear to occur in dustier TDE hosts with a high statistical significance. Third, we model two TDEs with MOSFiT to show that it is very easy to apply models to this standard dataset. Fourth, and finally, we apply the \texttt{FLEET} transient classifier \citep[]{gomez_fleet_2020, gomez_identifying_2023} to the data in OTTER and find that FLEET successfully classifies $\sim69\%$ of the sample as TDEs.

In the future, we plan to continue to maintain and expand the OTTER infrastructure. First, our aim is to add data on other transient events including GRBs, SLSNe, and SNe. Second, we want to connect existing modeling softwares, like \texttt{MOSFiT} \citep[]{guillochon_mosfit_2018} and \texttt{redback} \citep[]{sarin_redback_2024}, to the OTTER API, enabling easy access to models. For now, we have added tutorials to the API documentation on the best way to apply these modeling softwares to OTTER datasets. Another current limitation of OTTER is that we do not include spectroscopic observations. We encourage the continued use of WISEREP \citep[]{2012PASP..124..668Y} for making published spectra available. 

Finally, we want OTTER to be a tool for everyone. We will continue with an open source development model for the infrastructure code hosted on GitHub (\url{https://github.com/astro-otter}), and encourage feedback and contribution from the community on the web application and API. Please feel free to raise an issue on GitHub (or even open a pull request on GitHub!) with suggested improvements.

\begin{acknowledgments}
We thank Edo Berger for useful discussion and feedback, particularly on the classification confidence flagging. We thank the staff at SciServer, namely Charles Glaser, Gerard Lemson, and Arik Mitschang, for their indispendable discussion and assistance deploying OTTER and making it accessible to the world. We thank Ryan Chornock, Iair Arcavi, Sara Faris, Margaret Shepherd, and Yvette Cendes for testing the OTTER web interface and providing feedback. We thank Megan Masterson and Muryel Guolo for taking the time to provide data from their respective population papers. We thank Mathieu Renzo for useful comments on the paper draft. We thank the anonymous referee for their insightful comments.

N.F. acknowledges support from the National Science Foundation Graduate Research Fellowship Program under Grant No. DGE-2137419. KDA and CTC acknowledge support provided by the NSF through award SOSPA9-007 from the NRAO and award
AST-2307668. KDA gratefully acknowledges support from the Alfred P. Sloan Foundation. This research was supported in part by grant NSF PHY-2309135 to the Kavli Institute for Theoretical Physics (KITP).

\added{This research makes use of the SciServer science platform (www.sciserver.org). SciServer is a collaborative research environment for large-scale data-driven science. It is being developed at, and administered by, the Institute for Data Intensive Engineering and Science at Johns Hopkins University. SciServer is funded by the National Science Foundation through the Data Infrastructure Building Blocks (DIBBs) program and others, as well as by the Alfred P. Sloan Foundation and the Gordon and Betty Moore Foundation. We acknowledge that this research makes use of the IDIES Data Center, a resource developed and operated by the Johns Hopkins University, Institute for Data Intensive Engineering and Science (IDIES).}

\end{acknowledgments}

\newcommand{\softwaredelim}{--- }
\software{
numpy \citep[]{harris_array_2020} \softwaredelim
matplotlib \citep[]{hunter_matplotlib_2007} \softwaredelim
plotly \citep[]{plotly} \softwaredelim
astropy \citep[]{astropy_collaboration_astropy_2013, astropy_collaboration_astropy_2018, astropy_collaboration_astropy_2022} \softwaredelim
pandas \citep[]{reback2020pandas} \softwaredelim
synphot \citep[]{stsci_development_team_synphot_2018} \softwaredelim
pyarango \softwaredelim
astroquery \citep[]{ginsburg_astroquery_2019} \softwaredelim
ads \softwaredelim
skypatrol \citep[]{shappee_man_2014, kochanek_all-sky_2017} \softwaredelim
fundamentals \citep[]{Young_fundamentals} \softwaredelim
astro-datalab \softwaredelim
sparclclient \citep[]{juneau_sparcl_2024} \softwaredelim
pydantic \citep[]{Colvin_Pydantic_2025} \softwaredelim
fastapi \citep[]{Ramirez_FastAPI} \softwaredelim
starlette \softwaredelim
NiceGUI \citep[]{Schindler_NiceGUI_Web-based_user_2025}\softwaredelim 
Docker \citep[]{merkel_docker_2014} \softwaredelim
Minikube \softwaredelim
Kubernetes \softwaredelim
MOSFiT \citep[]{mockler_weighing_2019} \softwaredelim
FLEET \citep[]{gomez_fleet_2020, gomez_identifying_2023}
}

\bibliography{main, zotero, references_for_table}{}
\bibliographystyle{aasjournal}

\appendix

\section{Metadata and References for each TDE Candidate}\label{app:meta}
The default name, coordinates, redshift and discovery date (i.e. the metadata) for all of the TDE candidates in OTTER is given in Table \ref{tab:refs}. The rightmost column gives all of the references associated with this transient event in OTTER, including metadata and photometry. The reference numbers are defined below \autoref{tab:refs}. 

\begin{longtable}{p{3cm}p{2.5cm}p{2.5cm}p{2cm}p{2cm}p{0.5cm}p{3cm}}
\caption{Default metadata information for the transients in OTTER} \label{tab:refs} \\
\toprule
Transient Name & RA [HMS] & Dec [DMS] & Redshift & Discovery Date [MJD] & C & References \\
\midrule
\endfirsthead
\caption[]{Default metadata information for the transients in OTTER} \\
\toprule
Transient Name & RA [HMS] & Dec [DMS] & Redshift & Discovery Date [MJD] & C & References \\
\midrule
\endhead
\midrule
\multicolumn{7}{r}{Continued on next page} \\
\midrule
\endfoot
\bottomrule
\endlastfoot
NGC 5905 & 15:15:23.32 & +55:31:01.59 & 0.01124 & 1990-07-11 & 3.1 & \hyperref[bib:1]{ 1 }, \hyperref[bib:6]{ 6 }, \hyperref[bib:43]{ 43 }, \hyperref[bib:74]{ 74 }, \hyperref[bib:98]{ 98 }, \hyperref[bib:100]{ 100 }, \hyperref[bib:202]{ 202 }, \hyperref[bib:351]{ 351 } \\
RX J1420+53 & 14:20:24.2 & +53:34:11 & 0.147 & 1990-07-11 & 1.0 & \hyperref[bib:4]{ 4 }, \hyperref[bib:74]{ 74 }, \hyperref[bib:98]{ 98 }, \hyperref[bib:100]{ 100 }, \hyperref[bib:351]{ 351 } \\
RX J1624+75 & 16:24:56.7 & +75:54:57.5 & 0.0636 & 1990-08-25 & 1.0 & \hyperref[bib:3]{ 3 }, \hyperref[bib:43]{ 43 }, \hyperref[bib:74]{ 74 }, \hyperref[bib:98]{ 98 }, \hyperref[bib:100]{ 100 }, \hyperref[bib:202]{ 202 }, \hyperref[bib:351]{ 351 } \\
RBS 1032 & 11:47:26.8 & +49:42:59 & 0.026 & 1990-11-05 & 1.0 & \hyperref[bib:54]{ 54 }, \hyperref[bib:74]{ 74 }, \hyperref[bib:98]{ 98 }, \hyperref[bib:100]{ 100 }, \hyperref[bib:351]{ 351 } \\
LEDA 095953 & 13:47:30.33 & -32:54:50.63 & 0.0366 & 1991-07-06 & 1.0 & \hyperref[bib:11]{ 11 }, \hyperref[bib:74]{ 74 }, \hyperref[bib:98]{ 98 }, \hyperref[bib:100]{ 100 }, \hyperref[bib:351]{ 351 } \\
RX J1242-11A & 12:42:38.55 & -11:19:20.8 & 0.05 & 1992-07-15 & 1.0 & \hyperref[bib:2]{ 2 }, \hyperref[bib:4]{ 4 }, \hyperref[bib:43]{ 43 }, \hyperref[bib:74]{ 74 }, \hyperref[bib:98]{ 98 }, \hyperref[bib:100]{ 100 }, \hyperref[bib:202]{ 202 }, \hyperref[bib:351]{ 351 } \\
FIRST J153350.8+272729 & 15:33:50.88 & +27:27:29.618 & 0.032430 & 1995-04-16 & 1.0 & \hyperref[bib:287]{ 287 } \\
WINGS J1348 & 13:48:49.9 & +26:35:57.49 & 0.062 & 1999-12-20 & 1.0 & \hyperref[bib:47]{ 47 }, \hyperref[bib:64]{ 64 }, \hyperref[bib:74]{ 74 }, \hyperref[bib:98]{ 98 }, \hyperref[bib:100]{ 100 }, \hyperref[bib:351]{ 351 } \\
\end{longtable}

\clearpage\newpage
\begin{longtable}{p{0.3\textwidth} p{0.3\textwidth} p{0.3\textwidth}}
\endfirsthead \endhead \endfoot \endlastfoot
{[1] \citet{1996A&A...309L..35B} \label{bib:1} } & {[2] \citet{1999A&A...349L..45K} \label{bib:2} } & {[3] \citet{1999A&A...350L..31G} \label{bib:3} } \\ 
{[4] \citet{2000A&A...362L..25G} \label{bib:4} } & {[5] \citet{2000A&AS..143....9W} \label{bib:5} } & {[6] \citet{2002RvMA...15...27K} \label{bib:6} } \\ 
{[7] \citet{2006ApJ...653L..25G} \label{bib:7} } & {[8] \citet{2007A&A...462L..49E} \label{bib:8} } & {[9] \citet{2008ApJ...676..944G} \label{bib:9} } \\ 
{[10] \citet{2008ApJ...678L..13K} \label{bib:10} } & {[11] \citet{2009A&A...495L...9C} \label{bib:11} } & {[12] \citet{2009ApJ...698.1367G} \label{bib:12} } \\ 
{[13] \citet{2009PASP..121.1395L} \label{bib:13} } & {[14] \citet{2010ATel.2544....1D} \label{bib:14} } & {[15] \citet{2010ApJ...721..777A} \label{bib:15} } \\ 
{[16] \citet{2010ApJ...722.1035M} \label{bib:16} } & {[17] \citet{2011A&A...535L...1B} \label{bib:17} } & {[18] \citet{2011ATel.3384....1K} \label{bib:18} } \\ 
{[19] \citet{2011ATel.3463....1K} \label{bib:19} } & {[20] \citet{2011ATel.3811....1R} \label{bib:20} } & {[21] \citet{2011ApJ...732L..12B} \label{bib:21} } \\ 
{[22] \citet{2011ApJ...735..106D} \label{bib:22} } & {[23] \citet{2011ApJ...737...87J} \label{bib:23} } & {[24] \citet{2011ApJ...738...52L} \label{bib:24} } \\
\end{longtable}

\onecolumngrid
\clearpage
\section{Detailed Data Schema} \label{app:schema}

The details of the primary keys in each JSON file are (* means they are required):
\begin{enumerate}
    \item {\bf name*}: The name, and any aliases, of the transient. {\it alias} is used as a subkey to store a list of all of the possible names for the transient.
    \item {\bf coordinate*}: All reported coordinates for the transient, stored as a list.
    \item {\bf distance}: All reported distance measurements for this transient, stored as a list. Each value in the list has a {\it distance\_type} to describe the type of distance measurement stored. This {\it distance\_type} can be a redshift, luminosity distance, comoving distance, or dispersion measurement.
    \item {\bf date\_reference}: All relevant dates that have been reported for this transient, stored as a list. Each value has a {\it date\_type} keyword which currently can be the discovery date, peak date, or explosion date. These can either be observed or modeled parameters which will be denoted by the boolean {\it computed} keyword (true if it is a model parameter, false if it is observed).
    \item {\bf classification}: All reported classifications of this transient stored as a list. Every classification has a confidence associated with it which is derived analytically based on the information available in OTTER. This is described in more detail in \S\ref{sec:class-conf}.
    \item {\bf host}: Host information of reported galaxies associated with this transient. For each of these we store basic information like the host name, host coordinates, and host redshift. In the API, if there is no host associated, \texttt{astro-ghost} to attempt to find the host galaxy\citep[]{gagliano_ghost_2021}.
    \item {\bf photometry}: Photometry\footnote{Which is used broadly here to be any flux, flux density, or counts measurement at any frequency/wavelength} is a list organized into three primary categories based on the frequency/wavelength of the observation and stored in the {\it obs\_type} keyword: \texttt{xray}, \texttt{uvoir}, \texttt{radio}. Each photometric observation is required to have a flux/flux density and astropy-string units on that flux/flux density; a date and astropy date format; and filter name, filter effective frequency/wavelength, and units on that effective frequency/wavelength. If any corrections (host subtracted, k-correction, Milky Way extinction corrected, etc.) were applied to the dataset they are also required. If the data is given in units of counts, the telescope name is required so the telescope area is known and counts can be converted to a flux or flux density. These requirements are in place to ensure that {\it anyone} at {\it anytime} can compute the original photometric values and correct or convert them as they see fit.    
\end{enumerate}

More details on each of these primary keys, along with example JSON file snippets, are given in the subsequent subsections.

\subsection{Definitions}\label{sec:definitions}

The data for each transient object will be stored in individual \json\ files with a uniform structure, where the name of each file will be the same as the name of the transient object. Here, we describe the schema for the database, as well as the different types of data that can be stored in the \json\ files.

\begin{itemize}
    \item \textbf{object}: each astronomical transient object (TDE, SN, FRB, etc.) is referred to as an ``object". The default will be the IAU name of the object, followed by the shortest name alias.
    \item \textbf{\property{property}}: An overarching property of the object can be anything from ``photometry", a ``date", a ``classification", a ``host", etc. Can contain elements or subproperties. 
    \item \textbf{\subproperty{subproperty}}: any additional property within a property, useful when a large number of elements will share an additional keyword.
    \item \textbf{\keyword{keyword}}: the name of an item within a category, property, subproperty, or element in the {\tt json} dictionary. These keywords are dependent on the \property{property} or \subproperty{subproperty} that they belong to. Some commonly used examples are:
    \begin{enumerate}
        \item value (\S\ref{subsec:value}): the value of the element (e.g. ``17.2" for a magnitude).
        \item reference (\S\ref{subsec:reference}): the origin of the element, such as a paper, broker, catalog, etc.
        \item units (\S\ref{subsec:units}): units of the value written in astropy format (e.g. {\tt "km / s"})
        \item flag (\S\ref{subsec:flag}): integer to reference known flags.
        \item computed (\S\ref{subsec:computed}): a Boolean that specifies whether the value of the element was computed by the database (True), as opposed to an external reference (False). If not present it is assumed to be False.
        \item uui (\S\ref{subsec:uui}): universally unique identifier (UUI) for the element to be referenced elsewhere.
        \item default (\S\ref{subsec:default}): a Boolean used for when there are multiple entries of the element.
    \end{enumerate}
\end{itemize}

\subsection{Best practices} \label{sec:practices}

\begin{itemize}
    \item Use {\tt snake\_case} for naming variables, a naming convention where each word is in lower case and is separated by underscores.
    \item Use prefixes when properties or elements should be categorized together.
    \item Refrain from using plurals or capital letters when naming variables.
    \item Refrain from using keywords already used in the schema as names of properties or elements.
\end{itemize}

\subsection{Common Keywords} \label{sec:elements}

Here we provide a more detailed description and how to handle each of the optional keywords in an element, introduced in \S\ref{sec:definitions}.

\subsubsection{\keyword{value}} \label{subsec:value}

The most critical keyword in the database, which stores the value of any element, be that a parameter, measurement, or anything else. Can be a string, float, integer, or Boolean.

\subsubsection{\keyword{reference}} \label{subsec:reference}

The reference source for the value in the element. Whenever possible, the reference should be specified in the format for this is a 19-character ADS Bibcode (e.g. {\tt 2019ApJ...871..102N}). Only in the case of naming sources, if a bibcode is not available (e.g., it was automatically names by a telescope), a string or common source can also be input here (e.g. ``ZTF''). If {\tt computed = True}, the value of this keyword can be set to the uui of the original element used to compute the value, or multiple values (e.g. magnitude and redshift). In the json file, these can be either a string or a list of strings (in case there are multiple references for one value). A complete list of acceptable references is below:
\begin{enumerate}
    \item An ADS bibcode
    \item A uui corresponding to another item in the \json\ file.
    \item ``TNS" (Only in the case of name sources)
    \item ``Pan-STARRS" (Only in the case of name sources)
    \item ``GaiaAlerts" (Only in the case of name sources)
    \item ``ATLAS" (Only in the case of name sources)
    \item ``ZTF" (Only in the case of name sources)
    \item ``ASAS-SN" (Only in the case of name sources)
    \item ``WISeREP" (Only in the case of name sources)
    \item ``SOUSA" (Only in the case of name sources)
\end{enumerate}

\subsubsection{\keyword{units}} \label{subsec:units}

The units of the value. These should be formatted in astropy units format. For a description of this method \href{https://docs.astropy.org/en/stable/units/format.html#converting-from-strings}{see the Astropy documentation}. For a full list of acceptable units \href{https://docs.astropy.org/en/stable/units/#module-astropy.units.si}{see the list from Astropy}. Examples might include things like {\tt "km /s"},  {\tt "mag(AB)"} (for AB magnitudes), or {\tt "erg / ( s  cm2  Hz)"}.

\subsubsection{\keyword{flag}} \label{subsec:flag}

For situations that are too nuanced or rare to store in their own keywords, there is the option to store this information in the form of a flag. This can be an integer value for a flag associated with the value or element, or a comma separated list of flags. For a list of flags and their definitions see \S\ref{sec:flags}.

\subsubsection{\keyword{computed}} \label{subsec:computed}

A Boolean that defines whether the value is original from a reference (False), or if the value was computed using the database (True). The version of the database used to compute the values with {\tt computed == True} is stored in the {\tt version} element of the metadata.

\subsubsection{\keyword{uui}} \label{subsec:uui}

A universally unique identifier (UUI) used to cross-reference different elements with each other. These can be created with the {\tt uuid} Python package and look like this: {\tt c2ee78a2-acc0-4c01-b878-543130587e9a}.



\subsubsection{\keyword{default}} \label{subsec:default}

A Boolean value used in the event there are multiple competing entries for an important property such as the name of a transient, distance, or coordinates. There can only be one element with {\tt default == True} for each element in a property.

\subsubsection{\keyword{comment}} \label{subsec:comment}

An optional simple string with a comment pertaining to the individual element.

\subsection{Flags} \label{sec:flags}

There might be some situations that are relatively common, but either too complex to store as a keyword in an element, too long, or not necessary. In that case, there is the option to specify a flag. Additional flags can be added without having to re-define the entire schema. Multiple flags should be comma separated.

\subsection{Required Keywords} \label{sec:required}
This is a list of the required keywords for each type of data. Any other keywords listed in this can be assumed to be optional.

\subsubsection{Objects} \label{subsec:object_requirements}
At a minimum, each object is required to have the \keyword{name}, \keyword{coordinate}, and \keyword{reference\_alias} keywords. We require a reference for everything to ensure a high quality of data.

\subsubsection{Properties} \label{subsec:param_requirements}
The required values for all properties are:
\begin{itemize}
    \item \property{name}: Default name of the object. This is always required.
    \item \property{coordinate}: \keyword{equitorial} is always required with a \keyword{reference}. \keyword{galactic} and \keyword{ecliptic} are optional but must contain a \keyword{reference}.
    \item \property{distance}: Nothing required, ideally \keyword{redshift} will be provided. If one is provided, a \keyword{reference} is required!
    \item \property{epoch}: Nothing required, will be derived from photometry if provided. If one is provided, a \keyword{reference} is required!
    \item \property{classification}: Nothing required, TNS will be queried. If provided, a \keyword{reference} is required!
\end{itemize}

\subsubsection{Measurements} \label{subsec:measurement_requirements}
Each photometry point must have at least the following:
\begin{itemize}
    \item \keyword{reference}
    \item \keyword{raw}
    \item \keyword{raw\_units}
    \item \keyword{filter}
    \item \keyword{filter\_eff}
    \item \keyword{filter\_eff\_units}
    \item \keyword{date}
    \item \keyword{date\_format}
\end{itemize}

Additionally, if any corrections were applied the values \textit{must} be supplied as well! Similarly, if the magnitude is only an upper limit please set \texttt{upperlimit=True}.


\subsection{Properties} \label{sec:parameter}

\subsubsection{\property{name}} \label{sec:name}
The name information of the object. 

{\noindent \bf \subproperty{default\_name}} \label{sec:default_name}

The default name of the object. If available, this will be the IAU name without any prefixed (e.g. ``2019qiz"). If the object does not have an IAU name, the alias that is most similar to the IAU standard will be used.

{\noindent \bf \subproperty{alias}} \label{sec:alias}

All known names and aliases of the object are stored in this list, the default name will be selected from this list and stored in the \keyword{name} keyword.

\begin{lstlisting}[
    style=json,
    label=lst:example-json]
"default_name": "2019qiz",
"alias": [{
          "value": "AT2019qiz",
          "reference": "TNS"
          },
          {
          "value": "Melisandre",
          "reference": "2021ApJ...908....4V"
          },
          {
          "value": "ZTF19abzrhgq",
          "reference": "ZTF"
          }]
\end{lstlisting}

\subsubsection{\property{coordinate}} \label{sec:coordinate}

List that contains the coordinates of the object in the same format as the original reference. The default will be equatorial coordinates in degrees, this is the value that will be read by the pipeline, or computed if not available. There is the option to have multiple element entries, if for example there are different coordinates from different observatories. The default coordinates used for calculations will be identified with the keyword {\tt default = True}. This is different (but can be derived from) the optional coordinates values stored in for example, individual photometry measurements.

The type of coordinate is stored in the \keyword{coord\_type} element and can be equatorial, galactic, or ecliptic. Examples of how an object's coordinates and associated keywords can be stored are shown below. The simplest coordinate example with RA and DEC in degrees, and an associated reference:
\begin{lstlisting}[
    style=json,
    label=lst:example-json]
{
  "ra": 32.12178,
  "dec": 45.21694,
  "epoch": "J2000",
  "frame": "ICRS",
  "coord_type": "equatorial",
  "ra_units":"deg",
  "dec_units":"deg",
  "reference": "2019ApJ...871..102N",
}
\end{lstlisting}

\noindent
A more complex example with equatorial coordinates in hour angle, and computed galactic coordinates. In this case the set of coordinates with a reference is used as opposed to the one without a reference. Then the galactic coordinates are calculated based on the {\tt default = True} value, and the {\tt uui} associated is added.
\begin{lstlisting}[
    style=json,
    label=lst:example-json]
"coordinate": [
    {
    "ra": "04:46:37.880",
    "dec": "-10:13:34.90",
    "ra_units": "hourangle",
    "dec_units": "deg",
    "coord_type": "equatorial",
    "reference": "TNS",
    "computed": false,
    "uui": "c2ee78a2-acc0-4c01-b878-543130587e9a",
    "default": true
    },
    {
    "ra": "04:46:37.77867",
    "dec": "-10:13:34.6800",
    "ra_units": "hourangle",
    "dec_units": "deg",
    "coord_type": "equatorial",
    "reference": "Fake",
    "flag": "2,3"
    },
    {
    "l": 207.876557,
    "b": -32.322864,
    "l_units": "deg",
    "b_units": "deg",
    "coord_type": "galactic",
    "computed": true,
    "reference": "c2ee78a2-acc0-4c01-b878-543130587e9a"
    }]
\end{lstlisting}

The keywords that can be used within a coordinate element are:
\begin{itemize}
    \item \keyword{ra}: right ascension (str, int, float)
    \item \keyword{dec}: declination (str, int, float)
    \item \keyword{l}: galactic longitude (str, int, float)
    \item \keyword{b}: galactic latitude (str, int, float)
    \item \keyword{lon}: longitude (str, int, float)
    \item \keyword{lat}: latitude (str, int, float)
    \item \keyword{\_units}: suffix added to any of \keyword{ra}, \keyword{dec}, \keyword{l}, \keyword{b}, \keyword{lon}, \keyword{lat} to specify the units of the coordinate.
    \item \keyword{\_error}: suffix added to any of \keyword{ra}, \keyword{dec}, \keyword{l}, \keyword{b}, \keyword{lon}, \keyword{lat} to specify the uncertaintiy in the coordinate.
    \item \keyword{epoch}: epoch (e.g. J2000, B1950)
    \item \keyword{frame}: coordinate frame (e.g., ICRS, FK5)
    \item \keyword{default}: Boolean. If multiple entries, use this value as the default (True)
\end{itemize}

Each element in the \property{coordinate} property can only have one pair of coordinates in the same frame, but multiple elements can be added. 

\subsubsection{\property{distance}} \label{sec:distance}

This property stores different values realted to the distance to an object and can be anything like a redshift, dispersion measure, luminosity distance, etc. These can be computed or measured, some examples listed below:

\begin{lstlisting}[
    style=json,
    label=lst:example-json]
    "distance": [
        {
            "value": 1.1 ,
            "reference": "2019ApJ...871..102N" ,
            "computed": False,
            "default": True,
            "uuid": "c2ee78a2-acc0-4c01-b878-543130587e9b",
            "distance_type": "redshift"
        },
        {
            "value": 0.9 ,
            "error": 0.1
            "reference": "2019ApJ...871..102N" ,
            "computed": False,
            "distance_type": "redshift"
        },
        {
            "value": 1 ,
            "unit": " pc " ,
            "cosmology": "Planck18",
            "reference": "c2ee78a2-acc0-4c01-b878-543130587e9b",
            "computed": True
            "distance_type": "luminosity"
        },
        {
            "value": 0.1,
            "reference": "2019ApJ...871..102N",
            "computed": False,
            "distance_type": "dispersion_measure"
        }]
\end{lstlisting}

Where the keys can be the following.
\begin{itemize}
    \item \keyword{value}: The luminosity distance (float)
    \item \keyword{unit}: The units on the luminosity distance (str, astropy units)
    \item \keyword{error}: The error on the luminosity distance (float, optional)
    \item \keyword{reference}: The reference alias (integer)
    \item \keyword{cosmology}: Which cosmology was used to calculate the distance (str).
    \item \keyword{computed}: True if the value was computed, False otherwise (boolean)
    \item \keyword{distance\_type}: The type of distance measure. Can be "redshift", "luminosity", "dispersion\_measure", etc.
\end{itemize}

\subsubsection{\property{date\_reference}} \label{sec:date_reference}

These are mostly computed, even if it is something very simple like ``first data point". But if it is an explosion time determined from a complex model in a paper, that is not computed and a reference can be added. If a value is computed, the reference should be a uuid pointing to another epoch that it was computed from. Example:

\begin{lstlisting}[
    style=json,
    label=lst:example-json]
    "date_reference": [
    {
        "value": 56123.2 ,
        "date_format": "MJD",
        "date_type": "explosion",
        "reference":  "c2ee78a2-acc0-4c02-b878-543130587e9b",
        "computed": True
    },
    {
        "value": 56356.5,
        "date_format": "MJD"
        "reference": "c2ee78a2-acc0-4c02-b878-543130587e9b",
        "computed": True,
        "date_type": "peak"
    },
    {
        "value": "123456",
        "date_format": "MJD",
        "reference": "c2ee78a2-acc0-4c02-b878-543130587e9b",
        "computed": True,
        "date_type": "discovery"
    }
    ]
\end{lstlisting}

Each element can have the following keys:
\begin{itemize}
    \item \keyword{value}: The date of discovery
    \item \keyword{date\_format}: The format of the date
    \item \keyword{date\_type}: The type of date this is. Examples are ``explosion", ``peak", ``discovery".
    \item \keyword{reference}: The alias for the reference
    \item \keyword{computed}: If it was computed (bool)
\end{itemize}

\subsubsection{\property{classification}} \label{sec:classification}

The classification property is a dictionary with meta-level flag keywords and then all of the classifications stored under the \keyword{value} keyword. The current meta-level flags are \keyword{spec\_classed}, \keyword{unambiguous}, and \keyword{discovery\_method}, and are discussed in detail in \S\ref{sec:class-conf}. In addition to the keywords available in \S\ref{sec:elements}, classification \keyword{value} elements elements can have: \keyword{object\_class}, \keyword{confidence} (which is a flag, as defined in \S\ref{sec:class-conf}), and \keyword{ml\_score}. An object can have multiple classifications.

\begin{itemize}
    \item \keyword{object\_class}: the common object classes (e.g. SN, TDE, SN Ia, FRB, etc.)
    \item \keyword{confidence}: Flag with values defined in Table \ref{tab:conf-flag}.
    \item \keyword{ml\_score}: The machine learning score corresponding to this classification.
\end{itemize}

\begin{lstlisting}[
    style=json,
    label=lst:example-json]
"classification": {
    "spec_classed": true,
    "unambiguous": true,
    "value": [
      {
        "object_class": "SN Ia",
        "confidence": 3,
        "reference": "2018MNRAS.476..261B",
        "default": True
      },
      {
        "object_class": "SN",
        "confidence": 3,
        "reference": "TNS",
        "default": False
      },
      {
        "object_class": "SN Ia",
        "confidence": 1,
        "reference": "2017ApJ...476...61C",
        "default": False
      }
    ]
}
\end{lstlisting}

\subsubsection{\property{photometry}} \label{sec:photometry}

The photometry property will store all UV, optical, IR, X-ray, and radio photometry. This is what will likely be the most extensive element of the catalog. For some objects photometric measurements will have many repeated keywords such as photometric system, telescope, or whether or not they have some correction applied.

Groups of related photometry can be stored in individual arrays like in the example shown below.
\begin{lstlisting}[
    style=json,
    label=lst:example-json]
"photometry": [
  {
    "telescope": "ZTF",
    "mag_system": "AB",
    "reference": "2019ApJ...871..102N",
    "flux": [5.0, 5.0, 5.0, 5.0],
    "filter": ['r', 'r', 'r', 'r']
  },
  {
    "telescope": "ASAS-SN",
    "mag_system": "Vega",
    "reference": "2018MNRAS.476..261B",
    "raw": [5.0, 5.0, 5.0],
    "raw_err": [1.0, 1.0, 1.0]
  },
  {
    "filter": "Clear",
    "raw": [1]
  }
]
\end{lstlisting}
The full list of possible keywords that can be associated with a photometric measurement are the same for the phot\_N subproperty or the magnitude element, they are all listed below.


A measured magnitude is the most common element in the database, these can have a number of associated keywords. Each magnitude element can only have one flux-associated value (i.e. a magnitude element cannot have an AB magnitude and a flux). In addition to the default keywords listed for every element in \S\ref{sec:elements}, the keywords that can be associated with a magnitude element are:

\begin{itemize}
    \item \keyword{raw} : raw flux value, can be magnitude (for optical, infrared, or UV), energy (for X-ray), or flux density (for radio). This should be the closest available value to that measured by the telescope, without corrections, usually as published in a paper.
    \item \keyword{raw\_err} : error of raw flux value. This will be a ``default" error value for this flux measurement. If asymmetric errors were given, this is the average of the upper and lower values. All other forms of uncertainty given in \keyword{raw\_err\_details} are added in quadrature and entered here. If no \keyword{raw\_err\_details} keyword is given, this is simply the most conservative uncertainty estimate given.
    \item \keyword{raw\_units}: astropy units of the raw flux value.
    \item \keyword{raw\_err\_detail}: A list of dictionaries containing error details. The optional keywords here are ``upper" (the positive uncertainty in asymmetric errors), ``lower" (the negative uncertainty in asymmetric errors), ``systematic"  (the systematic uncertainty given), ``statistical" (the statistical undertainty given), and ``iss" (The scintillation uncertainty given for radio observations). While gathering data, we assume that if a single uncertainty measurement is provided, without details, that it is just the statistical uncertainty on the flux density measurement (i.e. the RMS noise of the radio image).  If multiple uncertainty measurements are provided, we combine them in quadrature for the ``flux\_err" keyword and then report the individual uncertainties in the ``error\_details" keyword.
    \item \keyword{value} : flux with corrections applied, can be a magnitude, energy, counts or flux density. If counts provided, must provide \keyword{telescope\_area}.
    \item \keyword{value\_err}: error on the corrected flux value. Same details as \keyword{raw\_err}.
    \item \keyword{value\_units}: units of the corrected flux value, can be astropy units or magnitude system.
    \item \keyword{value\_err\_detail}: Same as the \keyword{raw\_err\_detail} but for the \keyword{value} keyword.
    \item \keyword{epoch\_zeropoint} : if date is epoch, what is the offset.
    \item \keyword{epoch\_redshift} : if date is epoch, what is the redshift.
    \item \keyword{filter\_key} : The unique filter key in the format described in the metadata (\S\ref{sec:filter_alias})
    \item \keyword{obs\_type} : The type of observation made. Can be (``uvoir", ``xray", or ``radio")
    \item \keyword{telescope\_area}: collecting area of the telescope. This must be provided if \keyword{flux
    \_units} is ``counts"
    \item \keyword{date} : time of measurement.
    \item \keyword{date\_format} : astropy date format of the date measurement (e.g. mjd, jd, iso, etc).
    \item \keyword{date\_err} : uncertainty in the date value.
    \item \keyword{date\_min} : If the flux point is a binned value, this is the minimum date of that bin.
    \item \keyword{date\_max} : If the flux point is a binned value, this is the maximum date of that bin.
    \item \keyword{ignore} : was the data ignored.
    \item \keyword{upperlimit} : Boolean, is the mag an upper limit?
    \item \keyword{sigma} : significance of upper limit.
    \item \keyword{sky} : sky brightness in the same units as mag.
    \item \keyword{telescope} : telescope that took the data.
    \item \keyword{instrument} : instrument on telescope that took the data.
    \item \keyword{phot\_type} : is the photometry PSF, Aperture, or synthetic.
    \item \keyword{exptime} : Exposure time.
    \item \keyword{aperture} : If aperture photometry, aperture diameter in arcseconds.
    \item \keyword{observer} : Person or group that observed the data.
    \item \keyword{reducer} : Person who reduced the data.
    \item \keyword{pipeline} : Pipeline used to reduce the data.
    \item \keyword{corr\_k} : Boolean. Is the raw value k-corrected? Can be None which means that we are uncertain if it is k-corrected.
    \item \keyword{corr\_s} : Boolean. Is the raw value s-corrected? Can be None which means that we are uncertain if it is s-corrected.
    \item \keyword{corr\_av} : Boolean. Is the raw value Milky Way extinction corrected? Can be None which means that we are uncertain if it is av-corrected.
    \item \keyword{corr\_host} : Boolean. Is the raw value host subtracted? Can be None which means that we are uncertain if it is host subtracted.
    \item \keyword{corr\_hostav} : Boolean. Is the raw value corrected for intrinsic host extinction?  Can be None which means that we are uncertain if it is host extinction corrected.
    \item \keyword{val\_k} : Float. Value of the k-correction applied to mag.
    \item \keyword{val\_s} : Float. Value of the s-corrected applied to mag.
    \item \keyword{val\_av} : Float. Value of the Milky Way extinction correction applied to mag.
    \item \keyword{val\_host} : Float. Value of the host contribution applied to mag.
    \item \keyword{val\_hostav} : Float. Value of the intrinsic host extinction applied to mag.
    \item \keyword{xray\_model} : list of dictionaries of spectral modeling parameters used for calculating the X-ray flux for the photometry point. There should be one dictionary per X-ray photometry point. Each dictionary has the following required keywords: \keyword{model\_name}, \keyword{param\_names}, \keyword{param\_values}, \keyword{param\_units}, \keyword{min\_energy}, \keyword{max\_energy}, \keyword{energy\_units}. The dictionaries also support the following optional keywords: \keyword{param\_value\_upper\_err}, \keyword{param\_value\_lower\_err}, \keyword{param\_upperlimit}, \keyword{param\_descriptions}, \keyword{model\_reference} (which, unlike other references, can be either a bibcode, doi, or link to code documentation). All of the \keyword{param\_*} keywords should be lists of the same length but will differ in length from photometry point to photometry point depending on the spectral model used to obtain that value. An example: In the case of a powerlaw + blackbody model the param\_names could be something like [Gamma\_X, F\_PL, kT\_BB, F\_BB, R\_BB]. The other parameter keywords would then also be a length of five but with the actual values/constraints for each of these parameter names. If instead a, paper only uses a power law model, the param\_names would be [Gamma\_X, F].  
\end{itemize}

Clearly most magnitude measurements will not have all of these parameters available, but the user has the option to include up to all of these, in addition to the filter-specific keywords listed in \S\ref{sec:filter_alias}. Although it is recommended to store the filter-specific keywords in the metadata to avoid repetition. The {\tt mag} and {\tt raw} keywords do not necessarily have to contain magnitudes, they can contain flux values. This is acceptable as long as it is specified in the units field.

In the case of the magnitude element, the computed keyword does not apply. In this case {\tt raw} is always assumed to be not computed, and {\tt mag} is always assumed to be computed, where the computation can be as simple as applying a total correction of 0 mags.

\subsubsection{\property{host}} \label{sec:host}

Information about the host galaxy where the object is located. This will be a list of dictionaries with the following possible keys:
\begin{itemize}
    \item \keyword{host\_ra}: The Right Ascension of the host galaxy
    \item \keyword{host\_dec}: the Declination of the host galaxy
    \item \keyword{host\_ra\_units}: The units of the RA
    \item \keyword{host\_dec\_units}: The units of the declination
    \item \keyword{host\_z}: Redshift of the host galaxy
    \item \keyword{host\_type}: The classification of the host (ex. spiral, elliptical, dwarf, AGN, etc.)
    \item \keyword{host\_name}: The name of the host galaxy
    \item \keyword{reference}: The reference for this host information
\end{itemize}

\subsection{metadata} \label{sec:metadata}

Metadata is information that does not necessarily fit into any of the existing categories, but is instead used to support the database structure. Some examples of metadata are listed in this section.

\subsubsection{\property{reference\_alias}} \label{sec:reference_details}

We plan to store the 19 digit ADS bibcode for every reference for ease of analysis. This will store a mapping between those bibcodes and human readable names for easy visualization.

\begin{lstlisting}[
    style=json,
    label=lst:example-json]
"reference_alias": [
  {
    "name": "2019ApJ...871..102N",
    "human_readable_name": "somename et al. (2019)"
  },
  {
    "name": "2020ApJ...874...22N",
    "human_readable_name": "aname et al. (2020)"
  },
]
\end{lstlisting}

The two keywords used in this element are \keyword{name} and \keyword{human\_readable\_name}, which store the name of the reference and the corresponding alias integer, respectively. 

\subsubsection{\property{filter\_alias}} \label{sec:filter_alias}

The basic information about a filter can be stored in the metadata of the {\tt json} file for quick reference, without needing to add it to each individual photometry measurement.

\begin{lstlisting}[
    style=json,
    label=lst:example-json]
"filter_alias": [
  {
    "filter_key": "PAN-STARRS_PS1.r",
    "filter_name": "r"
    "wave_eff": 6155.47,
    "wave_min": 5391.11,
    "wave_max": 7038.08,
    "zp": 3173.02,
    "wave_units": "AA",
    "zp_units": "Jy",
    "zp_system": "Vega",
  }
]
\end{lstlisting}

Individual measurements can just reference the {\tt filter\_key} instead of storing this repeated information in each element. The keywords supported in the {\tt filter\_reference} element are:

\begin{itemize}
    \item \keyword{filter\_key}: keyword shared between observations and this reference. The format is adopted from the SVO, but replacing the slash {\tt /} with an underscore {\tt \_} character. These must be unique to the \keyword{wave\_eff}/\keyword{freq\_eff} keywords.
    \item \keyword{filter\_name}: The name of the filter in a more human-readable way. This will be useful for splitting up observations across multiple observatories for plotting.
    \item \keyword{wave\_eff}: effective wavelength of filter.
    \item \keyword{wave\_min}: minimum wavelength of filter.
    \item \keyword{wave\_max}: maximum wavelength of filter.
    \item \keyword{freq\_eff}: effective frequency of the filters (for radio!).
    \item \keyword{freq\_min}: minimum frequency of the filters (for radio!).
    \item \keyword{freq\_max}: maximum frequency of the filters (for radio!).
    \item \keyword{zp}: filter zeropoint.
    \item \keyword{wave\_units}: units of wavelengths.
    \item \keyword{freq\_units}: units of frequency.
    \item \keyword{zp\_units}: units of zeropoint value.
    \item \keyword{zp\_system}: system of zeropoint value (e.g. ``Vega").
\end{itemize}

\subsubsection{\property{schema\_version}} \label{sec:otter}

Version of the schema used.
\begin{lstlisting}[
    style=json,
    label=lst:example-json]
"schema_version": {
  "value": 1.0,
  "comment": "cite us please"
}
\end{lstlisting}

\clearpage
\section{Non-cumulative TDE candidate Discovery Histogram}\label{app:non-cumulative-hist} 
\begin{figure*}
    \centering
    \includegraphics[width=\linewidth]{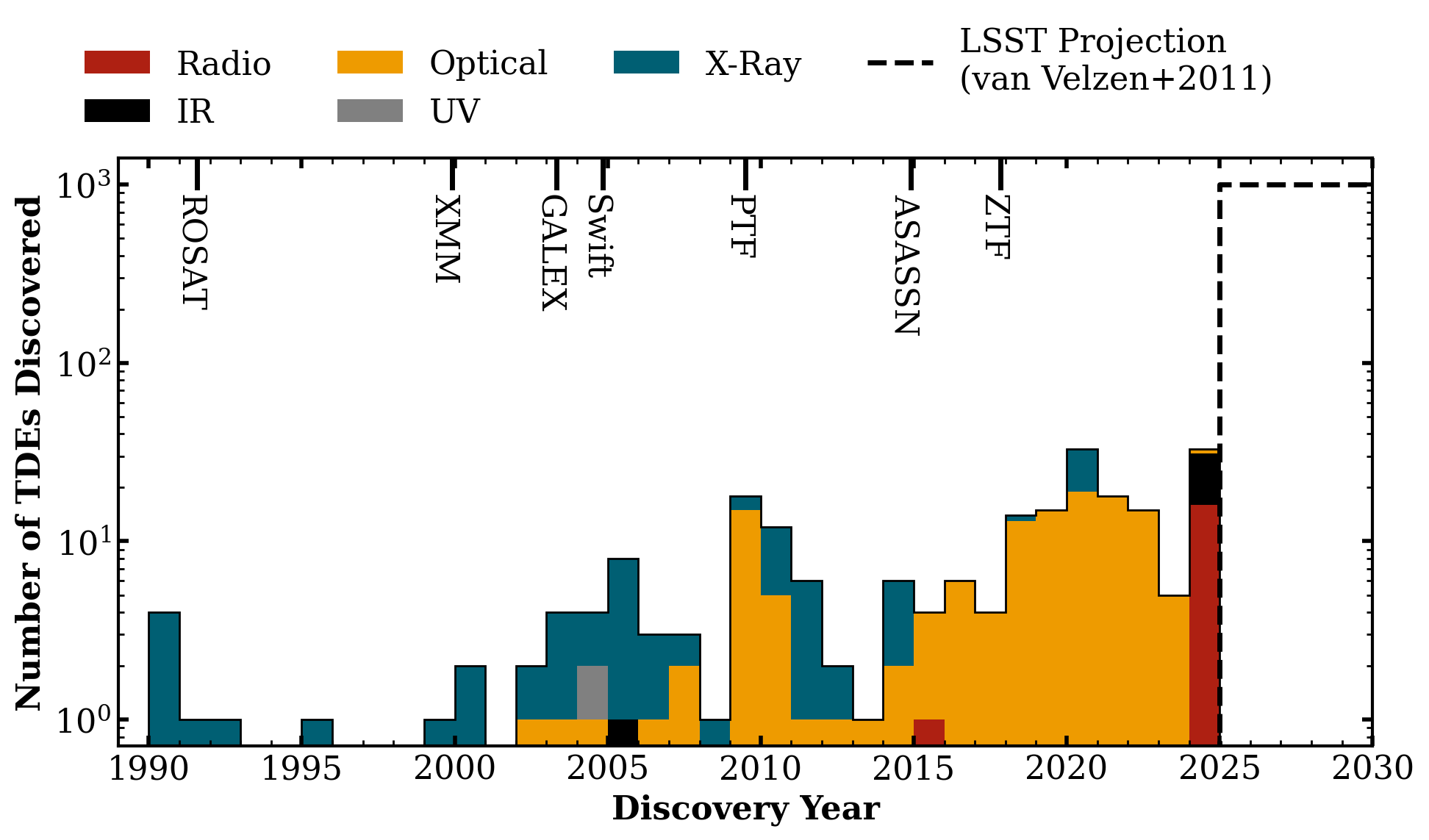}
    \caption{A stacked histogram of the number of TDE candidates discovered per year colored by discovery wavelength regime. Telescopes that contributed signficantly to the TDE candidate discovery rate are labelled along the top.}
    \label{fig:non-cumulative-disc-hist}
\end{figure*}

\autoref{fig:non-cumulative-disc-hist} shows a non-cumulative version of \autoref{fig:tde-discovery}. This is a different way to view the TDE candidate discovery rate.

\end{document}